\documentstyle[11pt]{article}

\renewcommand{\bibitem}{\item[]}
\newenvironment{biblist}
   {\begin{list}{}
    {\setlength{\leftmargin}{4ex}   %%% Move left margin right 4ex
     \setlength{\itemindent}{-4ex}  %%% move first line back left 4ex
     \setlength{\itemsep}{0ex}      %%% item separation
     \setlength{\parsep}{0ex}       %%% paragraph separation
   }}{\end{list}}

\newtheorem{theorem}{Theorem}[section]
\newtheorem{lemma}{Lemma}[section]
\newtheorem{definition}{Definition}[section]
\newtheorem{remark}{Remark}[section]

\newtheorem{prfT}{Proof of Theorem}[section]
\newenvironment{proofT}{\begin{prfT} \rm}{\hfill $\Box$ \end{prfT}}
\newtheorem{prfL}{Proof of Lemma}[section]
\newenvironment{proofL}{\begin{prfL} \rm}{\hfill $\Box$ \end{prfL}}
\newtheorem{prfC}{Proof of Corollary}[section]

\newtheorem{eg}{Example}[section]
\newenvironment{example}{\begin{eg} \rm}{\hfill $\Box$ \end{eg}}
\def \qed {\hfill $\Box$}
\begin{document}

Indian Statistical Institute Kolkata Tech. Rep. no. ASD/2010/3,
November 10, 2010

Revised draft August 1, 2011
\begin{center}{\Large{\textbf{Key Predistribution Schemes for Distributed Sensor Networks\\ via Block
Designs}}}
\\
\vskip.3in Mausumi Bose$^a$\footnote{Corresponding author.\\
\, \, \, email address:
mausumi.bose@gmail.com (Mausumi Bose)}, Aloke Dey$^b$, Rahul Mukerjee$^c$\\
\vskip.1in \emph{$^a$Indian Statistical Institute, Kolkata 700108,
India\\
 $^b$Indian
Statistical Institute, New Delhi 110016, India\\
 $^c$Indian Institute of Management Calcutta, Kolkata
700104, India}
\end{center}

\vskip.2in \noindent \textbf{Abstract} \, \,
 Key predistribution
schemes for distributed sensor networks have received significant
attention in the recent literature. In this paper we propose a  new
construction method for these schemes based on combinations of duals
of standard block designs. Our method is a broad spectrum one which
works for any intersection threshold. By varying the initial
designs, we can generate various schemes and this makes the method
quite flexible. We also obtain explicit algebraic expressions for
the metrics for local connectivity and resiliency. These schemes are
quite efficient with regard to connectivity and resiliency and at
the same time they allow a straightforward shared-key discovery.

%\vskip.1in \noindent \textbf{Keywords} \, \, association scheme, BIB
%design, connectivity, duals, PBIB design, resiliency.
%
%\vskip.1in \noindent \textbf{Mathematics Subject Classification} \,
%\,05B05, 68R05

\section{Introduction}\label{INT}

Distributed sensor networks have been extensively studied in recent
years due to their wide applicability in both civilian  and military
contexts. For instance, in a military operation, sensor nodes may be
distributed in a random manner over a sensitive area and, once
deployed, these nodes are required  to communicate with each other
in order to gather and relay information. This communication has to
be done in a secret manner and so secure keys need to be established
between the  nodes in the system. For more details on the
applications, the security framework and models for these
distributed sensor networks (DSNs) we refer e.g., to Carmen et al.
(2000), Roman et al. (2005) and Du et al. (2005). There are also
interesting results pertaining to an alternative situation where the
location of sensor nodes can be determined prior to deployment,
e.g., results by Younis et al. (2006), Martin et al., (2010),
Blackburn et al. (2010), Martin et al. (2011), and others. In this
paper we focus on the situation of random deployment of nodes.

Several authors have recommended the use of \emph{key
predistribution schemes} (KPSs) in a DSN, where secret keys are
installed in each sensor node before deployment. Eschenauer and
Gligor (2002) pioneered a probabilistic approach to key
predistribution and gave a scheme in which every node is assigned a
randomly chosen subset of keys from a given pool of keys. Chan et
al. (2003) generalized this basic scheme to the $q$-composite
scheme, where two nodes can  communicate \emph{only if} they share
at least $q$ common keys, where  $q$ is a prespecified integer
called the \emph{intersection threshold}.  Camtepe and Yener (2004)
first introduced the use of combinatorial designs in KPSs, using
finite projective planes and generalized quadrangles.  The principal
advantages of using deterministic key assignment schemes based on
combinatorial designs compared to random key assignment is that, in
the former approach, the problem of generating good pseudorandom
numbers is avoided, and moreover, by exploiting the combinatorial
structures of the underlying designs, one can study the local
connectivity and resiliency properties of the scheme easily, and
also carry out shared-key discovery and path-key establishment in a
structured manner. For more details on  these advantages we refer to
Lee and Stinson (2008) and Martin (2009).

Many researchers appreciated the advantages of the above approach
and continued to further develop this area. Lee and Stinson (2005a,
2005b) gave a construction based on transversal designs, Chakrabarti
et al. (2006) followed this by proposing a merger of a random
selection of blocks of a transversal design to form the nodes, Dong
et al. (2008) used $3$-designs, Ruj and Roy (2007) used partially
balanced designs and Ruj et al. (2009) used balanced incomplete
block designs in their construction. Lee and Stinson (2008) gave a
comprehensive account of key assignment schemes based on
combinatorial designs and studied all aspects of their schemes. They
 gave constructions for two classes of
 schemes, namely, a linear scheme with intersection threshold
$q=1$ and a quadratic scheme with $q=2$, based on transversal
designs.  They studied these two classes of schemes separately and,
for each of the two classes, they showed their scheme to be
efficient with regard to the levels of connectivity and resiliency,
while allowing simple shared-key discovery and path-key
establishment. The numbers of nodes required in the network for
these two classes of KPSs are of the form $p^2$ and $p^3$,
respectively, where $p$ is a prime or prime power.

In this paper we propose a new method for constructing KPSs and then
study the properties of the resulting schemes. Realizing  a
connection between the transversal designs used by Lee and Stinson
(2008) in their construction for $q=1$ and a particular type of
partially balanced incomplete block designs, we consider the latter
designs in their full generality and show that we can construct
useful KPSs based on a suitable combination of partially balanced
incomplete block designs. We propose one general construction method
for any given intersection threshold $q\; (\geq 1)$, and it will be
seen that for the case $q=1$, our construction covers the linear
scheme of Lee and Stinson (2008). One advantage of our proposed
method is that it works for all $q (\geq 1)$, and by varying the
choices of the designs, one can construct KPSs for networks with
varying numbers of nodes, key-pool sizes and numbers of keys per
node, thus providing more flexibility in choosing a scheme suitable
for the requirements of a situation. For example, now the number of
nodes need not be of the particular forms $p^2$ or $p^3$, with $p$
prime or prime power, as in Lee and Stinson (2008).
  These points will be elaborated on in Section~\ref{comparison}.

Another advantage of our method of construction is that it allows us
to obtain unified and explicit algebraic expressions for the metrics
for evaluating the connectivity and resiliency of these schemes, all
for general values of $q (\geq 1)$. Using these expressions, the
metrics can be easily calculated from the parameters of the
particular designs used in the construction.  This may be contrasted
with Lee and Stinson (2008), Ruj and Roy (2007) or Ruj et al.
(2009), where evaluation of the metrics can involve explicit
enumeration which may  become cumbersome.  We also show that our
KPSs have good connectivity with high levels of resiliency and the
combinatorial structure of the underlying designs make the
 shared-key discovery and path-key establishment phases
particularly simple.

In Section~\ref{preliminaries} of this paper we  give some
preliminaries on various metrics for evaluating a KPS, followed by
some basics on block designs.  Section~\ref{construction} describes
our proposed method for constructing a KPS. Next, in
Sections~\ref{connectivity} and \ref{resiliency} we obtain
expressions for the  connectivity and resiliency metrics for these
schemes and give illustrative examples. In
Section~\ref{applications} we apply our method to constructions
based on some specific block designs, together with numerical
illustrations. In Section~\ref{shared} we discuss how we can label
the keys and nodes so that shared-key discovery and path-key
establishment become simple. Finally in Section~\ref{comparison} we
discuss the gains achieved via our method of construction.

\section{Preliminaries}\label{preliminaries}
\subsection{Some metrics for evaluating KPSs}\label{metrics}
Several authors have considered some standard metrics for evaluating
the performance of key predistribution schemes for distributed
sensor networks. We briefly describe these metrics here; a more
comprehensive account can be found in Lee and Stinson (2008).

Two basic metrics of a KPS are the \emph{network size} or the number
of nodes in the network and the \emph{key storage} or the number of
keys stored per node, usually denoted by $n$ and $k$, respectively.
A KPS should typically have large $n$, say  1000 or much higher and
small $k,$ say about 50, though some authors have used $k$ up to
200.

 In a DSN the nodes are scattered over a physical area and, since nodes have
limited power, each can send or receive signals only over a certain
wireless communication range or \emph{neighborhood}. Once the nodes
are deployed, any two nodes which are within each other's
neighborhood can securely communicate directly with each other if
they have at least $q $ common keys, where $q (\geq 1) $ is a
specified integer,  the \emph{intersection threshold} of the DSN. On
the other hand, if two nodes in the same neighborhood do not have
$q$ common keys, then they can establish a connection through
multiple secure links  if there is a sequence of one or more
intermediate nodes connecting them such that every pair of adjacent
nodes in this sequence share $q$ common  keys.

To study the local connectivity of the network, we adopt the metrics
used in  Lee and Stinson (2005b, 2008), and for this, we now
introduce the relevant probabilities as defined by them. Define
${\rm Pr}_1$ to be the probability that two random nodes share at
least $q$ common keys. Thus given any two randomly chosen nodes
within each other's neighborhood, ${\rm Pr}_1$ is the probability
that these two nodes can establish secure direct communication with
each other. Also, define ${\rm Pr}_2$ to be the probability that two
 nodes in the same neighborhood do not have $q$ common keys but there
is a third node within the intersection of their neighborhoods which
shares $q$ common keys with both of them, thus allowing these two
nodes to communicate securely via this third node. So ${\rm Pr}_2$
is the probability that two randomly chosen nodes within the same
neighborhood fail to establish direct communication but can
communicate via a two-hop path.
 Hence, the sum  Pr = ${\rm Pr}_1+{\rm Pr}_2$ is a useful metric for studying
the local connectivity of a KPS through either a secure direct link
or a secure two-hop path.

Now suppose in an attack on the network a number of sensor nodes are
captured at random. Then it is assumed that all keys stored in these
compromised nodes are revealed and so cannot be used for
communication any more. Consider any two uncompromised nodes, say
$A$ and $A',$
 which have at least $q$ common keys. Then the
direct communication link between $A$ and $A'$ fails if  keys common
to them occur in one or more of the compromised nodes; otherwise,
the link remains secure. We want the sensor network to be resilient
against such random node compromises. From this consideration,
resiliency is measured by \emph{fail(s)}, which represents the
conditional probability of the link between $A$ and $A'$ to fail
when out of the remaining $n-2$ nodes, $s$ randomly chosen ones are
compromised, given that $A$ and $A'$ share at least $q$ common keys.
A smaller value of ${\rm fail}(s)$ implies a larger resiliency.

Finally, in order to communicate, two nodes in the same neighborhood
need to determine if they share $q$ common keys; this is the
\emph{shared-key discovery phase},
 and if they do not, then they try to
establish a secure two-hop path for communication; this is the
\emph{path-key establishment phase}. The difficulties involved in
these two phases are also used to assess the utility of a KPS.

\subsection{Some basics on block designs}\label{blockdesigns}

We present some basic definitions  of block designs and related
concepts which we will need in our constructions of KPSs.
Illustrative examples are also given. For more details on these
designs we refer to Street and Street (1987), Stinson (2003) and Dey
(2010).

\begin{definition}A {\it block design} $d^*$  is an arrangement of  a set of $v^*$ symbols
into $b^*$ subsets, these subsets being called {\it blocks}.
\end{definition}
\begin{example}\label{BIBDexample} The following is a block design $d^*$ with $v^*=9$, $b^*=12$. Denoting the symbols by $1, \ldots, 9$  and blocks by $1, \ldots,
12$, we can write
$$d^*: \begin{array}{cc||cc||cc||cc}
                                        Block  & Symbols &   Block& Symbols  & Block & Symbols & Block & Symbols\\
                                       1 & 4, 7, 2 & 4 & 5, 8, 3 & 7 & 6,9, 1 & 10& 1, 2, 3 \\
                                       2 & 7, 1, 5 & 5 & 8, 2, 6 & 8 & 9, 3, 4 & 11 & 4, 5, 6 \\
                                       3 & 1, 4, 8 & 6 & 2, 5, 9 & 9 & 3, 6, 7 & 12 & 7, 8, 9
 \end{array}$$\end{example}
\begin{definition}\label{dual} If $d^*$ is a block design with $v^*$ symbols and $b^*$ blocks then its {\it dual} design,
say $d$,  is a block design obtained from $d^*$ by interchanging the
roles of symbols and blocks, i.e., $d$ is a block design  involving
$b^*$ symbols and $v^*$ blocks, such that the $i$th block of $d$
contains the $j$th symbol if and only if the $j$th block of $d^*$
contains the $i$th symbol, $1\leq i\leq v^*, \, 1\leq j\leq b^*.$
\end{definition}

\begin{example}\label {BIBDdual} The dual design $d$
obtained from $d^*$ in Example~\ref{BIBDexample} has $12$ symbols,
$1. \ldots, 12$ and $9$ blocks denoted by $B_1, \ldots, B_{9}$ as
follows:
$$d:
\begin{array}{cc||cc||cc}
  Block  & Symbol &  Block & Symbol & Block & Symbol \\
  B_1 & 2, 3, 7, 10 & B_4 & 1, 3, 8, 11 & B_7 & 1, 2, 9, 12 \\
  B_2 & 1, 5, 6, 10 & B_5 & 2, 4, 6, 11 & B_8 & 3, 4, 5, 12 \\
  B_3 & 4, 8, 9, 10 & B_6 & 5, 7, 9, 11 & B_9 & 6, 7, 8, 12
\end{array}$$\end{example}

\begin{definition}\label{BIBD} A balanced
incomplete block  (BIB) design is a block design $d^*$ satisfying
the following conditions: (i) each symbol appears at most once in a
block, (ii) each block has a fixed number of  symbols, say $k^*$,
(iii) each symbol appears in a fixed number of blocks, say $r^*$,
and (iv) every pair of distinct symbols appear together in $\lambda$
blocks.
\end{definition}
The integer $\lambda$ is called the concurrence parameter of the BIB
design. It can be checked that the design in
Example~\ref{BIBDexample} is a BIB design with $\lambda = 1.$

\begin{definition}\label{Assoc} A relationship  defined on a set of symbols is called an association scheme with two associate classes
if it satisfies the following conditions:  (a) any two distinct
symbols are called either 1st or 2nd associates of each other, any
symbol being called the 0th associate of itself, (b) each symbol has
$\theta_j$ $j$th associates ($j=0,1,2$), and (c) for every pair of
symbols which are $j$th associates of each other, there are
$\phi^j_{u,w}$ symbols that are $u$th associates of one and $w$th
associates of the other $(j, u, w =0,1,2)$.
\end{definition}
The following  relations are evident from Definition~\ref{Assoc}:
%\sum_{j=1}^m\theta(j)&=&v-1;\;\;\;
\begin{equation}\label{relations}
\theta_0=1, \,\,
\phi^1_{0,0}=\phi^1_{0,2}=\phi^1_{2,0}=\phi^2_{0,0}=
\phi^2_{1,0}=\phi^2_{0,1}=0,\,\,
\phi^1_{0,1}=\phi^1_{1,0}=\phi^2_{0,2}=\phi^2_{2,0}=1.
\end{equation}

Various association schemes are available in the literature and for
these we refer to Clatworthy (1973). Our construction and results
are valid for any general association scheme but in our
illustrations in Section~\ref{applications}, we use three of these
association schemes, namely  group divisible, triangular  and Latin
square type association schemes. These are defined below.
\begin{definition}\label{gdscheme}
Let there be  $af$ symbols, ($a,f \geq 2)$, partitioned into  $a$
groups of $f$ symbols each, and let  the symbols in the $i$th group
be denoted by  $i1, i2, \ldots, if$, $i=1, \ldots, a$.  A group
divisible (GD) association scheme on these $af$ symbols is defined
as one where two distinct symbols are called $1$st associates if
they belong to the same group, and $2$nd associates otherwise.
\end{definition}
The above definition implies that for the GD association scheme, in
addition to  (\ref{relations}) we have  $\theta_1= f-1, \ \theta_2 =
f(a-1),$ $\phi^1_{1,1}=f-2, \ \phi^1_{1,2}=\phi^1_{2,1}=0, \
\phi^1_{2,2}=f(a-1), \ \phi^2_{1,1}=0, \
\phi^2_{1,2}=\phi^2_{2,1}=f-1, \ \phi^2_{2,2}=f(a-2).$

\begin{example}\label{gdscheme}
Let $a=2, f=3$. Then the $6$ symbols are partitioned into two groups
as: $\{11,  12,  13\}, \ \{21,  22 ,  23\}$. Now, for the symbol
$11$, the $1$st associates  are $12, 13$ while its $2$nd associates
 are $21, 22, 23.$
 Similarly, the $1$st and $2$nd associates of other symbols may be
written down and the parameters of the scheme can be obtained.
\end{example}
\begin{definition}\label{triangularscheme}
Let there be  $m \choose 2$ symbols, ($m \geq 4)$, denoted by
ordered pairs $ij$, $1\leq i<j \leq m$. A triangular association
scheme on these  symbols is defined as one where any two distinct
symbols are called $1$st associates if the ordered pairs
representing these symbols have one element in common, and $2$nd
associates otherwise.
\end{definition}
The above definition implies that for the triangular association
scheme, in addition to  (\ref{relations}) we have  $ \theta_1=
2(m-2), \theta_2 = {{m-2}\choose 2},$ $\phi^1_{1,1}=m-2,
\phi^1_{1,2}=\phi^1_{2,1}=m-3, \phi^1_{2,2}={{m-3}\choose 2},
\phi^2_{1,1}=4, \phi^2_{1,2}=\phi^2_{2,1}=2m-8,
\phi^2_{2,2}={{m-4}\choose 2}.$
\begin{example}\label{triangularscheme}
Let $m=5$. The $5\choose 2$ $(=10)$ symbols are denoted by the
ordered pairs: $12, 13, 14, 15$, $23, 24, 25, 34, 35, 45.$ Now, for
the symbol $12$, the $1$st associates  are $13, 14, 15, 23, 24, 25$
while its $2$nd associates are $34, 35, 45. $ Similarly, the $1$st
and $2$nd associates of other symbols may be written down and the
parameters of the scheme obtained.
\end{example}

\begin{definition}\label{Latinsquarescheme}
Let there be  $p^2$ symbols, $p\geq 3$, arranged in a $p\times p$
square $\mathcal{S}$ and suppose $k-2$ mutually orthogonal Latin
squares of order $p$ are available. A Latin square type association
scheme on these $p^2$ symbols is defined as one where any two
distinct symbols are called $2$nd associates if they occur in the
same row or same column of $\mathcal{S}$ or if, after superimposing
each of the Latin squares on $\mathcal{S}$, they occur in positions
occupied by the same letter in any of the Latin squares. Otherwise,
they are called $1$st associates.
\end{definition}
The above definition implies that for the Latin square type
 association scheme, in addition to  (\ref{relations}) we have  $
\theta_1= (p-1)(p-k+1), \ \theta_2 = k(p-1),$ $ \phi^1_{1,1}=
(p-k)(p-k-1)+p-2,\ \phi^1_{1,2}=\phi^1_{2,1}=k(p-k), \
\phi^1_{2,2}=k(k-1),\ \phi^2_{1,1}=(p-k)(p-k+1), \
\phi^2_{1,2}=\phi^2_{2,1}=(k-1)(p-k+1), \
\phi^2_{2,2}=(k-1)(k-2)+p-2.$

\begin{example}\label{Latinsquareschemeexample}
Let $p=4$ and $k=3.$  We denote the $4^2$ $(=16)$ symbols by the
ordered pairs: $11, 12, 13, 14, 21, 22, \ldots, 43, 44$ and write
$\mathcal{S }$ and the single Latin square $\mathcal{L}$ as
$$\mathcal{S} = \begin{array}{cccc}
                                                   11 & 12 & 13 & 14 \\
                                                   21 & 22 & 23 & 24 \\
                                                   31 & 32 & 33 & 34 \\
                                                   41 & 42 & 43 & 44
                                                 \end{array}, \hskip.5in \mathcal{L }=   \begin{array}{cccc}
                                                   A & B & C & D \\
                                                   B & C & D & A \\
                                                   C & D & A & B \\
                                                   D & A & B & C.
                                                 \end{array}
$$  Then it follows that for the symbol $11$, the $2$nd associates are $12, 13, 14, 21,
31, 41, 24, 33, 42,$ while its $1$st associates are $22, 23, 32, 34,
43, 44$. Similarly, the $1$st and $2$nd associates of other symbols
may be written down and the parameters of the scheme obtained.
\end{example}

\begin{definition}\label{PBIBD} Given an association scheme with two associate classes on a set of $v^*$ symbols,
a partially balanced incomplete block (PBIB) design based on this
association scheme is a block design $d^*$ with $v^*$ symbols and
$b^*$ blocks satisfying the following conditions:  (i)
 each symbol appears at most once in a block, (ii) each block has a
fixed number of symbols, say $k^*$, (iii) each symbol appears in a
fixed number of blocks, say $r^*$, and (iv) every pair of symbols
which are $j$th associates of each other appear together in
$\lambda_j$ blocks ($j=1,2$).
\end{definition}
The integers $\lambda_1$ and $\lambda_2$ are the two concurrence
parameters of the PBIB design, where $\lambda_1\neq \lambda_2$.

\begin{example}\label{gdexample}
We can construct a PBIB design $d^*$ based on the GD association
scheme by pairing each of the $af$ symbols with its second
associates to form the blocks. Thus, such a design can be
constructed for every integer $a, f (\geq 2)$. It is easy to see
that this design will have $v^*=af, \ b^*={a\choose 2}f^2, \ k^*=2,
\ r^*=(a-1)f$ and $\lambda_1=0,\ \lambda_2=1$. For example, a PBIB
design based on the GD association scheme in Example~\ref{gdscheme}
can be constructed by pairing each of the $6$ symbols with its
second associates to get $9$ blocks as follows:
$$d^* :
\begin{array}{cc||cc||cc}
           Block & Symbol & Block & Symbol & Block & Symbol \\
           1 & 11, 21 & 4 & 12, 21 & 7 & 13, 21 \\
           2 & 11, 22 & 5 & 12, 22 & 8 & 13, 22\\
           3 & 11, 23 & 6 & 12, 23 & 9 & 13, 23
         \end{array}.$$
  Clearly, this GD design
has $v^*=6, \ b^*=9,\  k^*=2, \ r^* = 3,\  \lambda_1=0,\
\lambda_2=1$.
\end{example}

\begin{example}\label{triangularexample}
We can construct a PBIB design $d^*$ based on the triangular
association scheme by pairing each of the $m\choose2$ symbols with
its second associates to get the blocks. Thus, such a design can be
constructed for every $m\geq 4$. It is easy to see that this design
will have $v^*= {m\choose 2}$, $ b^*=3{m\choose 4}$, $k^*=2$, $r^*=
{{m-2}\choose 2} $, and $\lambda_1=0,$  $\lambda_2=1.$  For example,
a PBIB design based on the triangular association scheme in
Example~\ref{triangularscheme}
 has $10$ symbols arranged in $15$
blocks given by: $(12, 34), (12, 35), (12, 45), (13, 24), (13, 25),
(13, 45), $ etc.
\end{example}

For a given positive integer $t(\geq 1)$, we now consider $t$ block
designs $d^*_1, \ldots ,d^*_t$ such that each $d^*_i$
 is a PBIB design based on an association scheme with two associate
classes and concurrence parameters $\lambda_1= 0$, $\lambda_2=1$,
the common occurrence number of  every symbol in  $d^*_i \ (i=1,
\ldots, t)$ being at least $t$. For $1\leq i\leq t$, consider the
dual $d_i$ of $d^*_i$ and denote the symbols of $d_i$ by
$1(i),\ldots ,v_i(i),$ and blocks by $B_1(i),\ldots ,B_{b_i}(i).$
Then from Definitions~\ref{dual} and \ref{PBIBD}, it is evident that
each such $d_i$, involving $v_i$ symbols and $b_i$ blocks, satisfies
the following conditions:

\noindent (I) every symbol occurs at most once in each block of
$d_i$,

\noindent (II) every symbol occurs in a fixed number of blocks, say
$r_i\;(2\leq
 r_i<b_i),$ of $d_i$,

\noindent (III)  every block of $d_i$ contains a fixed number of
symbols, say
 $k_i$ $\;(v_i>k_i\geq t)$, and

\noindent (IV) there is an association scheme  with two associate
classes \emph{on the set of blocks} of $d_i$; any two distinct
blocks either  have no common symbol, in which case they are called
1st associates of each other; or they have exactly one symbol in
common, in which case they are  called 2nd associates of each other;
every block being its own 0th associate.

For $1\leq i\leq t$, let $\theta_j(i)$  denote the number of $j$th
associates of any block of $d_i$, and given any two blocks which are
$j$th associates of each other, let $\phi^j_{u,w}(i)$ denote the
number of blocks of $d_i$ which are $u$th associates of one and
$w$th associates of the other ($j,u,w = 0,1,2).$ Then clearly, for
each design $d_i$ the relations corresponding to (\ref{relations})
hold, and moreover,
\begin{equation}\label{theta} \theta_0(i)=1,\;\;\;\;\theta_1(i)+\theta_2(i)=b_i-1 \ \ {\rm
and} \ \ \theta_1(i)>0, \  \ \theta_2(i)>0 \ (1\leq i\leq
t).\end{equation}

\begin{example}\label{gddual}
Let $d_1^*$ be the PBIB design  given in Example~\ref{gdexample}.
Then, the dual of $d_1^*$ is given by a design $d_1$ with $6$
symbols arranged in $9$ blocks. Denoting these symbols as $1(1),
\ldots, 9(1)$ and the blocks as $B_1(1), \ldots, B_{6}(1)$ as
described above, the design $d_1$ has blocks given by:
$$d_1: \begin{array}{cc||cc||cc}
  \mbox{Block} & \mbox{Symbols} & \mbox{Block} & \mbox{Symbols} & \mbox{Block} & \mbox{Symbols} \\
  B_1(1) & 1(1), 2(1), 3(1)  &   B_3(1) & 7(1), 8(1), 9(1) & B_5(1) & 2(1), 5(1), 8(1) \\
   B_2(1) & 4(1), 5(1), 6(1) &  B_4(1)  & 1(1), 4(1), 7(1) & B_6(1) & 3(1), 6(1), 9(1)
\end{array}$$
Clearly, $d_1$ satisfies conditions (I)-(III) above with $v_1 = 9, \
b_1 = 6,\  r_1=2, \ k_1 = 3$. Also, condition (IV) is satisfied; we
have the following association structure:
$$\begin{array}{c|c|c|}
   \mbox{Block}  & \mbox{1st associates} & \mbox{2nd associates} \\
    B_1(1)       &  B_{2}(1), B_{3}(1)   & B_{4}(1), B_{5}(1), B_{6}(1)  \\
    B_2(1)       &   B_{1}(1), B_{3}(1) &B_{4}(1), B_{5}(1), B_{6}(1)   \\
    B_3(1)       &   B_{1}(1), B_{2}(1)   & B_{4}(1), B_{5}(1), B_{6}(1) \\
    B_4(1)       & B_{5}(1), B_{6}(1)&  B_{1}(1),  B_{2}(1), B_{3}(1)\\
    B_{5}(1)     & B_{4}(1), B_{6}(1)&  B_{1}(1), B_{2}(1),    B_{3}(1)  \\
    B_6(1)       & B_{4}(1), B_{5}(1) & B_{1}(1), B_{2}(1),     B_{3}(1)
\end{array}$$
 So, in addition to the relations in (\ref{relations}), we have
$ \theta_1(1)=2, \ \theta_2(1)=3, \ \phi^1_{1,1}(1)=1,$ $
\phi^1_{1,2}(1)=\phi^1_{2,1}(1)=0, \ \phi^1_{2,2}(1)=3, \
\phi^2_{1,1}(1)=0, \ \phi^2_{1,2}(1)=\phi^2_{2,1}(1)=2, \
\phi^2_{2,2}(1)=0.$
\end{example}

 In the above development, we can as well take
any $d^*_i$ to be a BIB design with $\lambda=1, $ each symbol
appearing at least $t$ times in the design. Then by
Definitions~\ref{dual} and \ref{BIBD}, its dual design $d_i$ will
again satisfy the conditions (I)-(IV), but with $\theta_1(i)=0$.
This is because in this case, any two blocks of $d_i$ will always
have exactly one symbol in common and so by (IV), any two distinct
blocks of $d_i$ can only be second associates, there being no 1st
associates for any block. Thus, conditions (\ref{relations}) and
(\ref{theta}) are valid, keeping in mind that now in (\ref{theta}),
$\theta_1(i)=0$ and in (\ref{relations}), the quantities
$\phi^1_{u,w}(i)$ do not arise, while $\phi^0_{u,w}(i)=0$ and
$\phi^2_{u,w}(i)=0$ whenever $u=1$ or $w=1$.
\begin{example}\label{BIBDdualcontd}
Let $d_2^*$ be  the BIB design in Example~\ref{BIBDexample}. Then,
the dual of $d_2^*$ is the design  in Example~\ref{dual}, denoted by
$d_2$, say. Clearly, $d_2$ satisfies conditions (I)-(III) with $v_2
= 12, b_2 = 9, r_2=3, \ k_2 = 4$. Also, condition (IV) is satisfied
with no block in $d_2$ having any other block as its $1$st
associate, all distinct blocks being $2$nd associates of each other.
Thus, in addition to the relations in   (\ref{relations}), we have
$\theta_1(2)=0, \ \theta_2(2)=8, \ \phi^2_{1,1}(2)= \phi^2_{1,2}(2)=
\phi^2_{2,1}(2)=0,\ \phi^2_{2,2}(2)=7. $
\end{example}
In view of the above discussion, define two sets $Q$ and $\bar{Q}$
as
\begin{equation}\label{Q}
Q=\{i:1\leq i\leq t,\, \theta_1(i)>0\}\;\;{\rm
and}\;\;\bar{Q}=\{i:1\leq i\leq t,\, \theta_1(i)=0\}.
\end{equation}
Clearly, $i \in Q$ if  $d^*_i$ is a PBIB design and $i \in \bar{Q}$
if  $d^*_i$ is a BIB design as indicated above.

\section {Construction of KPS}\label{construction}

Suppose the intersection threshold of the required KPS is stipulated
as $q$.  We consider $t=q$ block designs $d_i^*, 1\leq i\leq t,$
where each $d_i^*$ is either a PBIB design with $\lambda_1=0,
\lambda_2=1$ or a BIB design with $\lambda=1$; every symbol
appearing at least $t$ times in each design. As before, for $1\leq
i\leq t,$ let $d_i$ be the dual of design $d^*_i,$ so  $d_i$
satisfies conditions (I)-(IV) listed  in
Subsection~\ref{blockdesigns}. A KPS with $q=t$, based on the
designs $d_1, \ldots, d_t$ is constructed as follows.

First identify the symbols in $d_1, \ldots, d_t$ as the keys of the
KPS. Next, consider all possible selections of one block from each
$d_i,\;1\leq i\leq t,$ and  take the union of the $t$ blocks in each
such selection as a node of the KPS. Thus the resulting KPS has $v=
\sum_{i=1}^t v_i$ keys given by the symbols $1(i),\ldots ,v_i(i),
(1\leq i\leq t)$ and $n = \Pi_{i=1}^t b_i$ nodes  given by
 \begin{equation}\label{nodes} N(\alpha_1\ldots \alpha_t)= B_{\alpha_1}(1)\cup \cdots \cup
B_{\alpha_t}(t),\;\;1\leq \alpha_i\leq b_i,\;\;1\leq i\leq
t.\end{equation}
 By condition
(III) in Subsection~\ref{blockdesigns}, every node has
$k=\sum_{i=1}^t k_i$ keys. Note that $n$ is multiplicative in the
$b_i$ while $k$ is additive in the $k_i,1\leq i\leq t$. As
illustrated later, this helps in attaining the twin objectives of
having a large number of nodes in the network  while keeping the
number of keys stored per node relatively small. \vskip.1in

\begin{remark}\label{covers1} {\rm One of the two constructions in Lee and Stinson
(2008), namely, the one with $q=1$, is covered by (\ref{nodes}).
 This fact will be elucidated in more detail in Remarks~\ref{covers2} and \ref{covers3}.}\qed \end{remark}

 For $1\leq i\leq t$, it is clear from (\ref{nodes})  that the block
$B_{\alpha_i}(i) $ is the contribution of the design  $d_i$ to the
node $N(\alpha_1\ldots \alpha_t)$. From this perspective, we
introduce the following definition.

\begin{definition}\label{projection} When nodes are constructed as in (\ref{nodes}), the
block of $d_i$ that appears in any node $A$ is called the projection
of the node $A$ on the design $d_i$ and is denoted by $proj(A,i)$.
\end{definition}

Thus from (\ref{nodes}), $B_{\alpha_i}(i)$ is the projection of the
node $N(\alpha_1\ldots \alpha_t)$ on $d_i$. We now define an
association scheme on the set of nodes as given by (\ref{nodes}).
This will play a crucial role in exploring the properties of the
KPSs obtained through (\ref{nodes}).  Here each associate
relationship is represented by a $t$-tuple of the form $j_1\ldots
j_t$.

\begin{definition}\label{assocnodes} Two distinct nodes $A$ and $A'$ are
$j_1\ldots j_t$th associates of each other if, for $1\leq i\leq t$,
$proj(A,i)$ and $proj(A',i)$ are $j_i$th associates of each other.
\end{definition}
\noindent We illustrate the above ideas with a small toy example
below.
\begin{example}\label{toy}
\emph{Toy Example:} Let $q=2$. So, by the above method, we take
$t=2$ and construct a KPS with $q=2$ based on two designs, $d^*_1$
and $d^*_2$. Let us take $d^*_1$ as the PBIB design  given in
Example~\ref{gdexample} and $d^*_2$ as the BIB design  in
Example~\ref{BIBDexample}. Their respective duals $d_1$ and $d_2$
are given in Examples~\ref{gddual} and \ref{BIBDdual}. The KPS
constructed by the above method has $n=b_1b_2=54$ nodes with
$k=k_1+k_2=7$ keys per node. Using (\ref{nodes}), we get the key
assignments in the nodes, for example, two  typical nodes are:

$N(1,1)=B_1(1)\cup B_1(2)= 1(1), 2(1), 3(1), 2(2), 3(2), 7(2),
10(2), $ and

$N(3,4)=B_3(1)\cup B_4(2)= 7(1), 8(1), 9(1), 1(2), 3(2), 8(2),
11(2).$

\noindent Then, by Definition~\ref{projection}, the blocks $B_1(1)$
and $B_1(2)$ are the projections of the node $N(1,1)$ on the designs
$d_1$ and $d_2$, respectively, i.e., $proj(N(1,1), 1)=B_1(1)$ and
$proj(N(1,1), 2)=B_1(2). $ Similarly, $proj(N(3,4), 1)=B_3(1)$ and
$proj(N(3,4), 2)=B_4(2).$ Now, from Examples~\ref{gddual} and
\ref{BIBDdualcontd}, we see that  $B_1(1)$ and $B_3(1)$ are $1$st
associates while $B_1(2)$ and $B_4(2)$ are $2$nd associates. So, by
Definition~\ref{assocnodes} we say that nodes $N(1,1)$ and $N(3,4)$
are $12$th associates of each other.
\end{example}

 In
Definition~\ref{assocnodes}, $j_1\ldots j_t\neq 0\ldots 0,$ since
the nodes $A$ and $A'$ are distinct. Also, by (\ref{Q}), $j_i= 0,
1\;{\rm or}\; 2$ if $i\in Q$ and $j_i= 0\;{\rm or}\; 2$ if $i\in
\bar{Q}$. Thus the set of all possible associate relationships
between two distinct nodes in the KPS is given by
%\footnotesize
{\begin{equation}\label {I} I=\{j_1\ldots j_t:j_1\ldots j_t\neq
0\ldots 0;\;j_i= 0, 1\;{\rm or}\; 2\;{\rm if}\;i\in Q\; {\rm and}\;
j_i= 0\;{\rm or}\;2\;{\rm if}\;i\in \bar{Q}\}.\end{equation}
%\normalsize

 We now obtain expressions for
certain parameters of the association scheme on the set of nodes, as
given by Definition~\ref{assocnodes}. For $j_1\ldots j_t\in I$, let
$n_{j_1\ldots j_t}$ denote the number of $j_1\ldots j_t$th
associates of any node $A$. Then by Definition~\ref{assocnodes},
$n_{j_1\ldots j_t}$ equals the product, over $1\leq i\leq t$, of the
number of $j_i$th associates of $proj(A,i)$. Therefore,
\begin{equation}\label{n} n_{j_1\ldots j_t}=\prod_{i=1}^t\theta_{j_i}(i).
\end{equation}
 Again, given
any two nodes which are $j_1\ldots j_t$th associates of each other,
let $p^{j_1\ldots j_t}_{u_1\ldots u_t,w_1\ldots w_t}$ denote the
number of nodes that are $u_1\ldots u_t$th associates of one node
and $w_1\ldots w_t$th associates of the other, where $j_1\ldots j_t,
u_1\ldots u_t$ and $w_1\ldots w_t \in I$.  Then as in (\ref{n}),
\begin{equation}\label{pij} p^{j_1\ldots j_t}_{u_1\ldots u_t,w_1\ldots
w_t}=\prod_{i=1}^t \phi^{j_i}_{u_i,w_i}(i).\end{equation}

Let $\lambda_{j_1\ldots j_t}$ denote the number of common keys
between any two distinct nodes $A$ and $A'$ which are $j_1\ldots
j_t$th associates of each other, $j_1\ldots j_t\in I $. Then from
Definition~\ref{assocnodes} it follows that
\begin{equation}\label{lambda}
\lambda_{j_1 \ldots j_t}=\sum_{i=1}^t\psi_{j_i}(i)
\end{equation} where $\psi_{j_i}(i)$ is the number of
symbols (or equivalently, keys) common to $proj(A,i)$ and
$proj(A',i)$ when they are $j_i$th associates of each other. By
condition (IV) of Subsection~\ref{blockdesigns} and the fact that
each block of $d_i$ is the 0th associate of itself, it is evident
that
\begin{equation}\label{psi}
 \psi_0(i)=k_i,\;\;\psi_1(i)=0,\;\;\psi_2(i)=1,\;\;1\leq i\leq t.
\end{equation}
We illustrate these concepts by continuing with the toy example in
Example~\ref{toy}.
\begin{example}\label{toycontd}
\emph{Toy Example continued:} Since $d^*_1$ is a PBIB and $d^*_2$ a
BIB design, by (\ref{I}), the set of all possible associate
relationships between any two nodes in the KPS is $I=\{ 02, 10, 12,
20, 22\}.$ Now, Examples~\ref{gddual} and \ref{BIBDdualcontd} show
that $\theta_1(1)=2, \ \theta_2(1)=3$ and $\theta_2(2)=8 $.
Recalling from (\ref{relations}) that $\theta_0(1)=\theta_0(2)=1$,
by (\ref{n}) it follows that the number of $02$th associates of any
node in the KPS is $n_{02}=1\times 8=8.$ Similarly, $n_{10}=2, \
n_{12}=16, \ n_{20}=3, \ n_{22}=24.$ Now, using the values of $
\phi^{j_1}_{u_1,w_1}(1)$ and $ \phi^{j_2}_{u_2,w_2}(2)$ from
Examples~\ref{gddual} and \ref{BIBDdualcontd} and remembering
(\ref{relations}), it follows from (\ref{pij}) that
$p^{12}_{02,10}=\phi^1_{01}(1)\phi^2_{20}(2)=1\times
1=1=p^{12}_{10,02},$ and similarly,
$p^{12}_{22,20}=p^{12}_{20,22}=3\times 1=3, \ p^{12}_{22,22}=3\times
7=21,\ p^{12}_{02,12}=p^{12}_{12,02}= 1\times 7=7, \
p^{12}_{10,12}=p^{12}_{12,10}= 1\times 1=1, \ p^{12}_{12,12}=
1\times 7=7,$ while every  other $p^{12}_{u_1u_2,w_1w_2}$ equals
zero.

Again, by (\ref{psi}), $\psi_0(1)=3,\ \psi_0(2)= 4, \ \psi_1(1)=0, \
\psi_2(1)= \psi_2(2)=1$, and so it follows from (\ref{lambda}) that
the number of symbols  common between any two nodes which are $02$th
associates of each other is $\lambda_{02}=3+1=4.$ Similarly,
$\lambda_{10}=4, \ \lambda_{12}=1, \lambda_{20}=5, \lambda_{22}=2.$
Hence, since $q=2$, all pairs of nodes, other than those which are
$12$th associates of each other, can communicate directly with one
another.
\end{example}

\section{Local connectivity}\label{connectivity}

 In this section we explore the  local connectivity of the KPS introduced
in (\ref{nodes}). Theorem~\ref{pr1pr2} is the main result in this
section and it gives an expression for the metric Pr for this
  scheme, in terms of the parameters of the constituent designs.
Some notation and two lemmas are needed in order to present the
theorem.  Let
\begin{equation} \label{delta} \Delta=\{j_1\ldots j_t:j_1\ldots j_t \in
I,\; \lambda_{j_1 \ldots j_t}\geq q\},\end{equation} where $I$ is
given by (\ref{I}). So, any two nodes which are $j_1\ldots j_t$th
associates of each other can communicate directly only if $j_1\ldots
j_t\in \Delta$. Let $\bar{\Delta}$ be the complement of $\Delta$ in
$I$ and let $\sum_{\Delta}$, $\sum_{\bar{\Delta}}$ and $\sum_I$
stand for sums over $j_1\ldots j_t\in \Delta$, $j_1\ldots j_t\in
\bar{\Delta}$ and $j_1\ldots j_t\in I$, respectively.

Given  two distinct nodes which are $j_1\ldots j_t$th associates of
each other, let $\mu_{j_1\ldots j_t}$ denote the number of nodes
sharing at least $q(=t)$ common keys with both of them. Also, for
any two distinct nodes $A$ and $A'$ in each other's neighborhood,
let the intersection of their neighborhoods contain $\eta$ nodes
excluding $A$ and $A'$ themselves. Define
\begin{equation}\label{beta}\beta_{j_1\ldots j_t}=1-\frac{{n-2-\mu_{j_1\ldots
j_t}\choose \eta}}{{n-2\choose \eta}},\;\;j_1\ldots j_t\in
\bar{\Delta}.\end{equation}

\begin{lemma}\label{lemma1} Any $j_1\ldots j_t \ (\in  I)$ is a member of $\Delta$ if
and only if either\newline (a) $j_i=0$ for at least one $i$, or (b)
$j_1=\cdots =j_t=2$.
\end{lemma}
\begin{proofL}
 Follows from (\ref {lambda}), (\ref{psi}) and  (\ref{delta}), noting that $k_i\geq t$ for
each $i$ by condition (III) of Subsection ~\ref{blockdesigns}.
\end{proofL}

 \begin{lemma}\label{lemma2} Given two distinct nodes which are  $j_1\ldots j_t$th associates
of each other,  if $j_1\ldots j_t\in \bar{\Delta}$, then $
\mu_{j_1\ldots j_t}=\sum\sum\;p^{j_1\ldots j_t}_{u_1\ldots u_t,
w_1\ldots w_t}\; , \ $ the double sum being  over $u_1\ldots u_t \in
\Delta$ and $w_1\ldots w_t \in \Delta$.
\end{lemma}
\begin{proofL} Follows from (\ref{delta}), on recalling the definition of
$p^{j_1\ldots j_t}_{u_1\ldots u_t,w_1\ldots w_t}.$
\end{proofL}

\begin{theorem}\label{pr1pr2}  The probability that  two distinct randomly chosen nodes
$A$ and $A'$ in each other's neighborhood can establish
communication, either directly or via a two-hop path, equals \, \,
\, \, ${\rm Pr}= {\rm Pr}_1 + {\rm Pr}_2$, where
\begin{equation}\label{pr1}{\rm Pr}_1=\frac{\sum_{\Delta}n_{j_1\ldots j_t}}{n-1}\; , \end{equation}
and \begin{equation}\label{pr2} {\rm
Pr}_2=\sum_{\bar{\Delta}}\frac{n_{j_1\ldots
j_t}}{n-1}\beta_{j_1\ldots j_t} \approx
\sum_{\bar{\Delta}}\;\frac{n_{j_1\ldots
j_t}}{n-1}\left[1-\left(1-\frac{\mu_{j_1\ldots
j_t}}{n-2}\right)^{\eta}\right ].
\end{equation}
\end{theorem}

%Change:
%\begin{theorem}\label{pr1pr2}  The probability that  two distinct randomly chosen nodes
%$A$ and $A'$ in each other's neighborhood can establish
%communication, either directly or via a two-hop path, equals \, \,
%\, \,  ${\rm Pr}= {\rm Pr}_1 + {\rm Pr}_2$, where
%\begin{equation}\label{pr1}{\rm Pr}_1=\frac{\sum_{\Delta}n_{j_1\ldots j_t}}{n-1}\; ,
%\\
%and \begin{equation}\label{pr2} {\rm
%Pr}_2=\sum_{\bar{\Delta}}\frac{n_{j_1\ldots
%j_t}}{n-1}\beta_{j_1\ldots j_t} \approx
%\sum_{\bar{\Delta}}\;\frac{n_{j_1\ldots
%j_t}}{n-1}\left[1-\left(1-\frac{\mu_{j_1\ldots
%j_t}}{n-2}\right)^{\eta}\right ].
%\end{equation}
%\end{theorem}

\begin{proofT} Let $C$ be the event that the nodes $A$ and $A'$ can
establish communication either directly or via a two-hop path.
Define $E(j_1\ldots j_t)$ as the event that $A$ and $A'$ are
$j_1\ldots j_t$th associates of each other. Since the events
$E(j_1\ldots j_t),\;j_1\ldots j_t \in I$, are mutually exclusive and
exhaustive, we can write
\begin{equation}\label{pr} {\rm
Pr}=P(C)=\sum_{I}P\{E(j_1\ldots j_t)\}P\{C|E(j_1\ldots
j_t)\},\end{equation} where $P\{C|E(j_1\ldots j_t)\}$ is, as usual,
the conditional probability of $C$, given $E(j_1\ldots j_t)$. Now,
for each $j_1\ldots j_t\in I$, recalling that there are
$n_{j_1\ldots j_t}$ nodes which are $j_1\ldots j_t$th associates of
any given node, it follows that
\begin{equation} \label{P(E)} P\{E(j_1\ldots j_t)\}=\frac{\frac{1}{2}n\times
n_{j_1\ldots j_t}}{{n\choose 2}}=\frac{n_{j_1\ldots
j_t}}{n-1}.\end{equation}

\noindent  Moreover, if $j_1\ldots j_t \in \Delta$, then by
(\ref{delta}), $A$ and $A'$ have at least $t$ common keys and hence
can establish direct communication, implying
\begin{equation}\label{P(E|C)} P\{C|E(j_1\ldots j_t)\}=1,\; {\rm for}\; j_1\ldots
j_t \in \Delta.\end{equation} On the other hand, if $j_1\ldots j_t
\in \bar{\Delta}$, then they have less than $t$ common keys. In this
case, direct communication between $A$ and $A'$ is not possible but
they can establish communication via a two-hop path provided the
intersection of their neighborhoods contains one of the
$\mu_{j_1\ldots j_t}$ nodes sharing at least $t$ common keys with
both of them.   Hence, using (\ref{beta}), it is clear that
\begin{equation}\label{P(C|E)} P\{C|E(j_1\ldots j_t)\}=\beta_{j_1\ldots j_t},\;
{\rm for}\;j_1\ldots j_t\in \bar{\Delta}.\end{equation} Substitution
of (\ref{P(E)}), (\ref{P(E|C)}) and  (\ref{P(C|E)}) in (\ref{pr})
establishes the theorem.
\end{proofT}

\begin{remark}\label{remark} {\rm The approximation used in (\ref{pr2}) is
 quite accurate when the quantities $n-2-\mu_{j_1,
\ldots, j_t} $ are large relative to $\eta$, which is typically the
case. Note also that the expression for ${\rm Pr}_2$ in (\ref{pr2})
is a refinement of that used in Lee and Stinson (2008) for $q=2$. To
see this, first note from (\ref{pr1}) that
\begin{equation}\label{1-pr1} \frac{\sum_{\bar{\Delta}}n_{j_1\ldots
j_t}}{n-1}=\frac{n-1-\sum_{\Delta}n_{j_1\ldots j_t}}{n-1}= 1-{\rm
Pr}_1,\end{equation} because $\sum_I n_{j_1\ldots j_t}=n-1$. Next,
write $\mu^*= {\rm min}\{\mu_{j_1\ldots j_t}:j_1\ldots j_t\in
\bar{\Delta}\}$ and from (\ref{beta}) observe that $\beta_{j_1\ldots
j_t}\geq \beta^*$ for every $j_1\ldots j_t\in \bar{\Delta}$, where
$\beta^*$ is defined as in (\ref{beta}) with $\mu_{j_1\ldots j_t}$
replaced by $\mu^*$. As a result, from (\ref{pr2}) and
(\ref{1-pr1}), we get
\begin{equation}\label{pr2bound}
{\rm Pr}_2\geq \sum_{\bar{\Delta}}\frac{n_{j_1\ldots
j_t}}{n-1}\beta^*=(1 - {\rm Pr}_1)\beta^* \approx (1 - {\rm
Pr}_1)\left[1-\left(1-\frac{\mu^*}{n-2}\right)^{\eta}\right].
\end{equation}
For their quadratic scheme, Lee and Stinson (2008) took ${\rm Pr}_2$
as the counterpart of the lower bound in (\ref{pr2bound}) for their
setup. Instead, we work here with the more direct expression given
in (\ref{pr2}), and in addition, this is valid for all $q\geq 1$.}
\qed
\end{remark}

\begin{remark}\label{algebraic} {\rm Lee and Stinson (2008)
 remarked that it is difficult to
find an algebraic expression of $\mu^*$ for their quadratic KPS, and
therefore, studied  ${\rm Pr}_2$ through design specific numerical
evaluation of $\mu^*$. An advantage of our method is that for all
 $q (\geq 1)$, even when one starts with arbitrary designs, Theorem~\ref{pr1pr2} gives readily applicable algebraic expressions for both ${\rm Pr}_1$ and
${\rm Pr}_2$ for our schemes in terms of the design parameters.
Equations (\ref{theta}), (\ref{n}), (\ref{pij}),  and Lemmas
~\ref{lemma1} and ~\ref{lemma2} can be used in finding the
$n_{j_1\ldots j_t}$ and $\mu_{j_1\ldots j_t}$, and hence one can
find ${\rm Pr}_1$ and ${\rm Pr}_2$  explicitly in specific
situations. The following examples serve to illustrate this point
for the cases $q=1$ and $q=2.$ } \qed
\end{remark}

\begin{example}\label{example1} Case: $q=1$. We take $t=1$ and construct a KPS
as in (\ref{nodes}) with $d^*_1$ either (a) a PBIB or (b) a BIB
design.

\noindent (a) If  $d^*_1$ is a PBIB design with $\lambda_1=0,
\lambda_2=1$, then  its dual design $d_1$ has $\theta_1(1)> 0$. Then
$n=b_1$ and by (\ref{Q}),
 (\ref{I}) and Lemma~\ref {lemma1}, $Q = \{1\}, \, I = \{1, 2\}, \, \Delta=
\{2\} \,\,{\rm and}\,\,\bar{\Delta}=\{1\}.$ Also, from (\ref{n}) and
(\ref {pij}),\, $n_1=\theta_1(1), \, n_2=\theta_2(1)$ and
$p^1_{2,2}=\phi^1_{2,2}(1)$. So by Lemma~\ref{lemma2},
$\mu_1=p^1_{2,2}=\phi^1_{2,2}(1)$. Hence (\ref{pr1}) and (\ref{pr2})
yield
\begin{equation}\label{prex1} {\rm Pr}_1=\frac{\theta_2(1)}{b_1-1}\;\;\;\; {\rm
and}\;\;{\rm Pr}_2\approx
\frac{\theta_1(1)}{b_1-1}\left[1-\left(1-\frac{\phi^1_{2,2}(1)}{b_1-2}\right)^\eta\right].\end{equation}

\noindent (b) If $d^*_1$ is a BIB design with $\lambda=1$, then its
dual $d_1$ has $\theta_1(1)=0$, $\theta_2(1)=b_1-1.$ Then $n=b_1$
and by (\ref{Q}), (\ref{I}) and Lemma~\ref {lemma1}, $\bar{Q} =
\{1\}, \, I = \{2\}= \Delta.$ So by (\ref{pr1}), $ {\rm Pr_1}
=\frac{b_1-1}{b_1-1}=1$ always.
\end{example}

 \begin{remark}\label{covers2} {\rm As mentioned in the Remark~\ref{covers1}, the
construction in Lee and Stinson (2008) with $q=1$ is covered by
(\ref{nodes}). To see this in detail,  we first note that in their
construction, the nodes are taken as the blocks of a transversal
design (cf. Stinson (2003)), with $kp$ symbols and $p^2$ blocks,
such that (a) the set of symbols is partitioned into $k$ groups each
of cardinality $p$, (b) each group contributes one symbol to each
block, and (c) any two symbols from different groups occur together
in exactly one block.
%It can be now be checked  that the
%blocks of the transversal design used in the construction in Lee and
%Stinson (2008) with $q=1$, are  subject to an association scheme
%with two associate classes -- any two distinct blocks are either
%first associates in which case they do not intersect, or second
%associates in which case they intersect in one symbol.

Recalling Definitions~\ref{Latinsquarescheme} and \ref{PBIBD} it can
now be checked that such a transversal design is actually the dual
of a PBIB design based on a Latin square type association scheme
with $v^*=p^2, b^*= kp, r^* = k, k^*=p, $  and $\lambda_1=0,
\lambda_2=1.$ Hence one can verify that their construction can
equivalently be described via our construction in (\ref{nodes}) with
$t =1$ and $d_1^*$ chosen as this PBIB design. Then its dual $d_1$
is their transversal design involving $v_1=kp$ symbols and $b_1=p^2$
blocks, such that conditions (I)--(IV) of
Subsection~\ref{blockdesigns} hold with $r_1= p, \, k_1= k, \,
\theta_1(1)=(p-1)(p+1-k), \, \theta_2(1)=k(p-1), \,
\phi^1_{2,2}(1)=k(k-1).$ Hence we can apply (\ref{prex1}) to  get
$${\rm Pr}_1=\frac{k}{p+1}\;\;\;{\rm and}\;\;\;{\rm Pr}_2\approx
\left(1-\frac{k}{p+1}\right)\left[1-\left(1-\frac{k(k-1)}{p^2-2}\right)^\eta\right].$$
These exactly match the expressions for ${\rm Pr}_1$ and  ${\rm
Pr}_2$ in Subsection 4.1.1 of Lee and Stinson (2008). We will see in
Remark~\ref{covers3} that their expression for ${\rm fail}(s)$ also
follow from our corresponding expressions. }\qed
\end{remark}
\begin{example}\label{toycontd2} Case: $q=2$. \emph{Toy example:} We continue with the KPS considered in Examples~\ref{toy} and
\ref{toycontd}. From the  $\lambda_{j_1j_2}$ values in
Example~\ref{toycontd}, it follows that $\Delta=\{02, 10, 20, 22\}$
and so, using the $n_{j_1j_2}$ values obtained there,
 (\ref{pr1}) gives ${\rm Pr}_1 = (8+2+3+24)/53 = 0.6981$. To obtain ${\rm Pr}_2$, we
 see that
 $\bar{\Delta}=\{12\}$, and so, remembering the values of $p^{12}_{u_iu_2,w_1w_2}, u_1u_2,w_1w_2\in\Delta,$ obtained in
 Example~\ref{toycontd}, it follows from Lemma~\ref{lemma2} that $\mu_{12} = 1+1+3+3+21=29.$
Hence, from (\ref{pr2}), ${\rm Pr}_2
=\frac{16}{53}[1-(1-29/52)^\eta]$ and for varying values of $\eta$
we have
$$\begin{tabular}{c|c|c|c|c|c|c|c|c}
  $\eta$ & 1 & 2 & 3 & 4 & 5 & 10 & 15 & 20 \\ \hline
 ${\rm Pr}_1+{\rm Pr}_2$ & 0.8665 & 0.9409  &  0.9739 & 0.9884  & 0.9949   & 0.9999 & 1.0000 & 1.0000 \\
\end{tabular}$$
\end{example}

\begin{example}\label{example2}
General Case, $q=2$: (a) PBIB and BIB design: Suppose we construct a
KPS as in (\ref{nodes}) based on two designs $d_1^*$ and $d_2^*$
given by a PBIB design with $\lambda_1=0, \lambda_2=1$ and a BIB
design with $\lambda=1$, respectively. Hence their duals $d_1$ and
$d_2$ have
 $\theta_1(1)> 0$ and $\theta_1(2)= 0$. Then $n=b_1b_2$
and by (\ref{Q}), (\ref{I}) and Lemma~\ref{lemma1}, we have  $Q
=\{1\}, \, \bar{Q}=\{2\}, \, I =\{02,10,12,20,22\}, \, \Delta=\{02,
10, 20, 22\} \, {\rm and} \, \bar{\Delta}=\{12\}.$ Also, by
(\ref{theta}) and (\ref{n}), $n_{02}=\theta_2(2), \,
n_{10}=\theta_1(1), \, n_{12}=\theta_1(1)\theta_2(2), \
n_{20}=\theta_2(1)\, {\rm and } \, n_{22}=\theta_2(1)\theta_2(2).$
So from (\ref{pr1}), on using (\ref{theta}), we have
\begin{eqnarray}\label{pr1ex2}
{\rm Pr_1} & =&
\frac{1}{b_1b_2-1}\{\theta_2(2)+\theta_1(1)+\theta_2(1)+\theta_2(1)\theta_2(2)\}\,
, {\nonumber}\\[1ex]
& =& \frac{1}{b_1b_2-1}\{b_1+b_2-2+\theta_2(1)\theta_2(2)\}\,.
\end{eqnarray}

Next by (\ref{pij}) and Lemma~\ref{lemma2},
\begin{eqnarray}
\mu_{12}&=&\sum\sum
p^{12}_{u_1u_2,w_1w_2}=\sum\sum\phi^1_{u_1,w_1}(1)\phi^2_{u_2,w_2}(2)\,
 \nonumber \\[2ex]
&=&\phi^1_{0,0}(1)\phi^2_{2,2}(2)+\phi^1_{0,1}(1)\phi^2_{2,0}(2)+\phi^1_{0,2}(1)\phi^2_{2,0}(2)+\phi^1_{0,2}(1)\phi^2_{2,2}(2)\,
 {\nonumber}\\[2ex]
&&+\phi^1_{1,0}(1)\phi^2_{0,2}(2)+\phi^1_{1,1}(1)\phi^2_{0,0}(2)+\phi^1_{1,2}(1)\phi^2_{0,0}(2)+\phi^1_{1,2}(1)\phi^2_{0,2}(2)\,
 {\nonumber}\\[2ex]
&&+\phi^1_{2,0}(1)\phi^2_{0,2}(2)+\phi^1_{2,1}(1)\phi^2_{0,0}(2)+\phi^1_{2,2}(1)\phi^2_{0,0}(2)+\phi^1_{2,2}(1)\phi^2_{0,2}(2)\,
 {\nonumber}\\[2ex]
&&+\phi^1_{2,0}(1)\phi^2_{2,2}(2)+\phi^1_{2,1}(1)\phi^2_{2,0}(2)+\phi^1_{2,2}(1)\phi^2_{2,0}(2)+\phi^1_{2,2}(1)\phi^2_{2,2}(2).\nonumber
\end{eqnarray}
 Hence invoking (\ref{relations}) for the association schemes underlying the designs $d_1$ and $d_2$, we get
\begin{equation}\label{muex2}\mu_{12}=2+2\phi^1_{1,2}(1)+2\phi^1_{2,2}(1)
+\phi^1_{2,2}(1)\phi^2_{2,2}(2).\end{equation}
 Since
$\bar{\Delta}=\{12\}$ and $n_{12}=\theta_1(1)\theta_2(2)$,
(\ref{pr2}) now yields
\begin{equation}\label{pr2ex2}
 {\rm Pr}_2\approx
\frac{\theta_1(1)\theta_2(2)}{b_1b_2-1}\left[1-\left(1-\frac{\mu_{12}}{b_1b_2-2}\right)^\eta\right],\end{equation}
with $\mu_{12}$ as given in (\ref{muex2}).
\end{example}

\begin{example}\label{example3} General Case $q=2$: (b) Both PBIB designs: Now suppose we construct a KPS as in (\ref{nodes})
based on two PBIB designs, each with $\lambda_1=0$ and $
\lambda_2=1$, resulting in $\theta_1(1)$ and $\theta_1(2)$ both
positive. Then $n=b_1b_2$ and by (\ref{Q}), (\ref{I}) and
Lemma~\ref{lemma1}, $Q=\{1, 2\}, I = \{01, 02, 10, 11, 12, 20, 21,
22\}$, $\Delta=\{01, 02, 10, 20, 22\}$ and $\bar{\Delta}=\{11, 12,
21\}$. Hence proceeding as in Example~\ref{example2}, one can check
that
\begin{eqnarray}
{\rm Pr}_1&=&\frac{1}{b_1b_2-1}\{b_1+b_2-2+\theta_2(1)\theta_2(2)\},{\nonumber}\\[1ex]
n_{11}&=&\theta_1(1)\theta_1(2), \, n_{12}=\theta_1(1)\theta_2(2), \, n_{21}=\theta_2(1)\theta_1(2){\nonumber}\\[1ex]
\mu_{11}&=&2+\phi^1_{2,2}(1)\phi^1_{2,2}(2),\, \, \, \mu_{12}=2+2\phi^1_{1,2}(1)+2\phi^1_{2,2}(1)+\phi^1_{2,2}(1)\phi^2_{2,2}(2),{\nonumber}\\[1ex]
\mu_{21}&=&2+2\phi^1_{1,2}(2)+2\phi^1_{2,2}(2)+\phi^2_{2,2}(1)\phi^1_{2,2}(2).{\nonumber}
\end{eqnarray}${\rm Pr}_2$ can be readily obtained  using these expressions
for the $n_{j_1j_2}$ and $\mu_{j_1j_2},$ $j_1j_2\in \bar{\Delta}$,
in (\ref{pr2}).
\end{example}

\section{Resiliency}\label{resiliency}

 We now study the resiliency of the KPS as given by (\ref{nodes}) and for this we recall the notion of
${\rm fail}(s)$ introduced in Subsection~\ref{metrics}.
 Theorem~\ref{fails} below gives an algebraic expression for
${\rm fail}(s)$ and it is the main result of this section. Some
notation and a lemma are needed in order to present the theorem.

 Let $A$ and
$A'$ be two distinct nodes which have at least $t$ common keys,
i.e., by (\ref{delta}), they are $j_1\ldots j_t$th associates of
each other, for some $j_1\ldots j_t \in \Delta.$ Then by
Lemma~\ref{lemma1}, the set $\Omega = \{i: 1\leq i \leq t, \ j_i=0
\;\; {\rm or } \; \;2\}$ is nonempty. For $i \in \Omega$, let
$\delta_{j_i}(i)$ equal $1$ or $r_i$ according as $j_i= 0$ or $2$,
respectively. Consider now any nonempty subset $\Gamma$ of $\Omega$.
Then for $i\in \Gamma$,  as noted in (\ref{psi}),  $proj(A,i)$ and
$proj(A',i)$ are identical if $j_i=0$, while $proj(A,i)$ and
$proj(A',i)$ have exactly one common key if $j_i=2$. Define $H(A,A';
\Gamma)$ as the collection of nodes $A''$, such that for every $i\in
\Gamma$,  $proj(A'',i)$ is different from $proj(A,i)[=proj(A',i)]$
whenever $j_i=0$, and $proj(A'',i)$ does not include the single key
common to $proj(A,i)$ and $proj(A',i)$ whenever $j_i=2$.

\begin{lemma}\label{lemma3}
 With reference to any two distinct nodes $A$ and $A'$
which are $j_1\ldots j_t$th associates of each other, where
$j_1\ldots j_t \in \Delta,$  the cardinality of $H(A,A'; \Gamma)$
defined as above is given by
$$\sigma(\Gamma) = \left(\prod_{i \in \Gamma} \{b_i-\delta_{j_i}(i)\}\right)\left(\prod_{i \notin \Gamma} b_i
\right)
 .$$

\end{lemma}
\begin{proofL} In view of the definition of the $\delta_{j_i}(i)$, this is evident from (\ref{nodes}) on recalling that every symbol occurs
in $r_i$ blocks of $d_i$ by condition (II) of
Subsection~\ref{blockdesigns}.
\end{proofL}

\begin{theorem}\label{fails}  Let $\xi_{j_1\ldots
j_t}=\Pi_{i=1}^t\xi_{j_i}(i)$, where
$$\xi_0(i)=1-(1-b_i^{-1})^s,\;\;\xi_1(i)=1,\;\;\xi_2(i)=1-(1-r_ib_i^{-1})^s, \ \ 1 \leq i \leq t.$$
Then for     $s<{\rm min}(k_1,\ldots ,k_t),$ $${\rm fail}(s)\approx
1-\left(\frac{n}{n-2}\right)^s+\left(\frac{n}{n-2}\right)^s
\frac{\sum_{\Delta}n_{j_1\ldots j_t}\xi_{j_1\ldots
j_t}}{\sum_{\Delta}n_{j_1\ldots j_t}}.$$

\end{theorem}
\begin{proofT}Consider two distinct nodes $A$ and $A'$. Let $D$ denote the
event that they have at least $q(=t)$ common keys and $F$ denote the
event that the link between them fails when out of the remaining
$n-2$ nodes, $s$ randomly chosen ones are compromised. Then
\begin{equation}\label{P(F|D)}
 {\rm fail}(s)=P(F|D)=P(F\cap D)/P(D).\end{equation}
As in the proof of Theorem~\ref{pr1pr2}, let $E(j_1\ldots j_t)$
denote the event that $A$ and $A'$ are $j_1\ldots j_t$th associates
of each other. Then by (\ref{delta}) and (\ref{P(E)}),
\begin{equation}\label{P(D)}
P(D) =\sum_{\Delta}P\{E(j_1\ldots
j_t)\}=\frac{\sum_{\Delta}n_{j_1\ldots j_t}}{n-1}.\end{equation}
Similarly,\begin{eqnarray}\label{P(FcapD)}
P(F\cap D)&=&\sum_{\Delta}P\{F\cap E(j_1\ldots j_t)\}\nonumber \\
&=&\sum_{\Delta}P\{E(j_1\ldots j_t)\}P\{F|E(j_1\ldots j_t)\}\nonumber \\
&=&\sum_{\Delta}\frac{n_{j_1\ldots j_t}}{n-1}P\{F|E(j_1\ldots
j_t)\}.\end{eqnarray}

In order to find an expression for the conditional probability in
(\ref{P(FcapD)}), take any fixed $j_1\ldots j_t\in \Delta$, and
condition on the event that $A$ and $A'$ are $j_1\ldots j_t$th
associates of each other. Then as  noted in the context of
Lemma~\ref{lemma3}, the set $\Omega=\{i:1\leq i\leq t,j_i=0\; {\rm
or}\; 2\}$ is nonempty. By (\ref{psi}), $proj(A,i)$ and $proj(A',i)$
have one or more common keys \emph{if and only if} $i\in \Omega$.
For any such $i$, let $G_i$ denote the event that not all of the
key(s) common to $proj(A,i)$ and $proj(A',i)$ occur in one or more
of the $s$ randomly chosen nodes that are compromised. Then for the
fixed $j_1\ldots j_t$ under consideration, by the usual union
intersection formula,
\begin{equation}\label{P(F|E)}
 P\{F|E(j_1\ldots j_t)\}=1 -P\{\cup_{i\in
\Omega}\;G_i\}=1+\sum_{\Gamma \subseteq
\Omega}(-1)^{|\Gamma|}P(\cap_{i\in \Gamma}\;G_i),\end{equation}
where the sum on the extreme right is over all nonempty subsets
$\Gamma$ of $\Omega$, and $|\Gamma|$ denotes the cardinality of
$\Gamma$. Note that the right side of (\ref{P(F|E)}) depends on
$j_1\ldots j_t$ through $\Omega$.

For any fixed nonempty subset $\Gamma$ of $\Omega$, we now find the
probability $P(\cap_{i\in \Gamma}\;G_i)$ appearing in
(\ref{P(F|E)}). Denote the $s$ randomly chosen nodes that are
compromised by $A_1^*,\ldots ,A_s^*$. Fix any $i\in \Gamma$, so that
$j_i= 0$ or 2. First suppose $j_i= 0$. Then $proj(A,i)$ and
$proj(A',i)$ are identical, and $G_i$ happens \emph{if and only if},
for each $1\leq l\leq s$, $proj(A_l^*,i)$ is different from
$proj(A,i)[=proj(A',i)]$. The \emph{only if} part of this claim is
obvious. The \emph{if} part follows because any two distinct blocks
of $d_i$ intersect in at most one symbol or key ({\it vide}
condition (IV) of Subsection~\ref{blockdesigns}) and $s<{\rm
min}(k_1,\ldots ,k_t)$. Next, let $j_i= 2$. Then $proj(A,i)$ and
$proj(A',i)$ have exactly one common key and $G_i$ happens \emph{if
and only if}, for each $1\leq l\leq s$, $proj(A_l^*,i)$ does not
include this single common key. Recalling the definition of
$H(A,A';\Gamma)$, it is now clear that $\cap_{i\in \Gamma}\;G_i$
happens \emph{if and only if }each of $A_1^*,\ldots ,A_s^*$ belongs
to $H(A,A';\Gamma)$. So, as $n=\prod_{i=1}^t b_i$, by
Lemma~\ref{lemma3}, we get
\begin{eqnarray}\label{P(capGi)}
 P(\cap_{i\in \Gamma}\;G_i)&=&\frac{{\sigma(\Gamma) \choose s}}{{n-2 \choose s}} \approx \left(\frac{\sigma(\Gamma)}{n-2}
\right)^s \nonumber \\
&=&\left(\frac{n}{n-2}\right)^s\left(\frac{\sigma(\Gamma)}{n}\right)^s\nonumber \\
&=&\left(\frac{n}{n-2}\right)^s\prod_{i\in
\Gamma}\left(1-\frac{\delta_{j_i}(i)}{b_i}\right)^s.
\end{eqnarray}

Since $\xi_{j_i}(i)=1$ for $j_i=1$, i.e., for $i\notin \Omega$, and
$$1-\left(1-\frac{\delta_{j_i}(i)}{b_i}\right)^s=\xi_{j_i}(i),$$ for $j_i=0$ or 2, i.e., for $i\in \Omega$, substitution of (\ref{P(capGi)})
in (\ref{P(F|E)}) yields
\begin{eqnarray}\label{P(F|E)2nd}
P\{F|E(j_1\ldots j_t)\}&\approx &1+\left(\frac{n}{n-2}\right)^s\sum_{\Gamma \subseteq \Omega}\;(-1)^{|\Gamma|}\prod_{i\in \Gamma}\left(1-\frac{\delta_{j_i}(i)}{b_i}\right)^s \nonumber \\
&=&1-\left(\frac{n}{n-2}\right)^s+\left(\frac{n}{n-2}\right)^s\prod_{i\in \Omega}\left[1-\left(1-\frac{\delta_{j_i}(i)}{b_i}\right)^s\right]\nonumber \\
&=&1-\left(\frac{n}{n-2}\right)^s+\left(\frac{n}{n-2}\right)^s\prod_{i=1}^t\xi_{j_i}(i)\nonumber \\
&=&1-\left(\frac{n}{n-2}\right)^s+\left(\frac{n}{n-2}\right)^s\xi_{j_1\ldots
j_t}. \end{eqnarray} If we now substitute (\ref{P(F|E)2nd}) in
(\ref{P(FcapD)}) and then substitute (\ref{P(D)}) and
(\ref{P(FcapD)}) in (\ref{P(F|D)}) the result follows.
\end{proofT}

\begin{remark} {\rm The approximation in (\ref{P(capGi)}) and hence that in
Theorem~\ref{fails} is  in the spirit of Lee and Stinson (2008). It
is quite accurate when $n$ and $\sigma(\Gamma)$ are large and $s$ is
relatively small, which is typically the case.}\qed
\end{remark}
\begin{remark} {\rm The condition $s<{\rm min}(k_1,\ldots ,k_t)$ in Theorem~\ref{fails}
is not severe because typically $s$ is not large. Moreover, it can
be checked that for the case $q=t =1$, Theorem~\ref{fails} remains
valid even without this condition.}\qed
\end{remark}
Examples~\ref{example1} and \ref{example2} are now revisited with a
view to illustrating Theorem~\ref{fails}. Example~\ref{example3} can
also be treated in the same way as Example~\ref{example2} and so is
not shown here.

\begin{example}\label{example1contd} Example~\ref{example1}
(continued).   Here $t =1, \, n=b_1$ and, irrespective of whether
$d^*_1$ is a PBIB design with $\lambda_1=0, \ \lambda_2=1$, or a BIB
design with $\lambda=1$, we have  $\Delta=\{2\}$. Hence
Theorem~\ref{fails} yields
\begin{equation}\label{failsex5.1}
{\rm fail}(s)\approx
1-\left(\frac{n}{n-2}\right)^s+\left(\frac{n}{n-2}\right)^s\xi_2(1)=1-\left(\frac{b_1-r_1}{b_1-2}\right)^s.
\end{equation}\end{example}

\begin{remark}\label{covers3} {\rm As a continuation of Remarks~\ref{covers1} and \ref{covers2},
we now see that the ${\rm fail}(s)$ values of the linear scheme
constructed in Lee and Stinson (2008) also follow from
Thoerem~\ref{fails}. Since their scheme  has $b_1=p^2$ and $r_1= p$,
on substituting these in our expression (\ref{failsex5.1}) we get
$${\rm fail}(s)\approx 1-\left(\frac{p^2-p}{p^2-2}\right)^s. $$ This
matches the expression for ${\rm fail}(s)$ in their Subsection
4.1.1.} \qed
\end{remark}

\begin{example}\label{example2contd} Example~\ref{example2}
(continued). Here $t =2,\, \theta_1(1)>0,\, \theta_1(2)= 0,\,
n=b_1b_2$ and $\Delta=\{02, 10, 20, 22\}$. As noted earlier,
\begin{equation}\label{nex2}
n_{02}=\theta_2(2),\,n_{10}=\theta_1(1),\,n_{20}=\theta_2(1),n_{22}=\theta_2(1)\theta_2(2).\end{equation}
Also, \begin{eqnarray}\label{xiex2}
\xi_{02}&=&\{1-(1-b_1^{-1})^s\}\{1-(1-r_2b_2^{-1})^s\},\nonumber\\
\xi_{10}&=&1-(1-b_2^{-1})^s,\nonumber\\
\xi_{20}&=&\{1-(1-r_1b_1^{-1})^s\}\{1-(1-b_2^{-1})^s\},\nonumber\\
\xi_{22}&=&\{1-(1-r_1b_1^{-1})^s\}\{1-(1-r_2b_2^{-1})^s\}.
\end{eqnarray} One can now readily apply Theorem~\ref{fails} to find fail$(s).$
\end{example}

\section{Applications}\label{applications}

As mentioned earlier, our method of construction, based on
(\ref{nodes}) and applicable to any $q (\geq 1)$, can yield KPSs for
widely diverse values of the underlying parameters such as the
number of nodes $n$, the number of keys per node $k$ and the key
pool size $v$, thus enabling the practitioner to find a suitable KPS
depending on the requirements of a given situation. This flexibility
arises because of the freedom in choosing the PBIB or BIB designs
$d^*_1, \ldots ,d^*_t$ that one starts with while applying
(\ref{nodes}). Furthermore, the analytical results in the last two
sections can be applied to ensure that the resulting KPSs behave
nicely with regard to local connectivity and resiliency, as measured
by Pr and fail$(s)$.

In order to give a flavor of the points noted above without making
the presentation too long, we now focus on the case $q = 2$ and in
the next three subsections present three applications where $d^*_1$
is a PBIB design based on the (a) GD, (b) triangular and (c) Latin
square type association schemes, and $d^*_2$ is a BIB design; note
that these correspond to the setup of Example~\ref{example2}. The
parameter values of the resulting KPSs, obtained via (a), (b) and
(c) are seen to be

(a) $n=af(2g+1)$, $k=(a-1)f+g$, $v={a\choose
2}f^2+\frac{1}{3}(2g+1)g$, where $a, f (\geq 2)$ are any integers
and $g (\geq 3)$ satisfies $g= 0\;{\rm or}\; 1\;({\rm mod}\; 3)$,

 (b) $n={m\choose 2}(2g+1)$, $k={{m-2}\choose 2}+g$,
$v=3{m\choose 4}+\frac{1}{3}(2g+1)g$, where $m (\geq 4)$ is any
integer and $g$ is as in (a),

 (c) $n = p^2(2g+1)$, $k = \tilde{k}+g $, $v = \tilde{k}p +
\frac{1}{3}(2g+1)g$, where $p (\geq 3)$ and $\tilde{k} (< p+1)$ are
integers such that $\tilde{k}-2$ mutually orthogonal Latin squares
of order $p$ exist, and $g $ is as in (a).

\noindent Thus these three applications alone are capable of
producing KPSs for a wide range of parameter values. Moreover,
Theorems~\ref{pr1pr2} and \ref{fails} allow us to explore the
properties of these KPSs and the examples in the next three
subsections show that they can behave quite well with respect to Pr
and fail$(s)$. Indeed, our construction in (\ref{nodes}), coupled
with these theorems, can easily allow numerous other choices of
$d^*_1$ and $d^*_2$ as well, and hence paves the way for obtaining
 KPSs with an even more versatile range of parameter
values, while ensuring attractive values for Pr and fail$(s)$. In
contrast, the existing methods of construction are almost invariably
design specific, i.e., they employ only BIB designs or only
transversal designs and so on, and as a result, it is very difficult
for these methods to achieve parameter values as diverse as what is
achieved, for instance, in (a)-(c) above. In addition, the existing
methods are not always informative about the properties of the
resulting KPSs with regard to local connectivity or resiliency. We
will return to this comparison in more detail in the concluding
section.

\subsection {Use of a PBIB design based on the group divisible association scheme and a BIB design
}\label{gd}

Suppose the design $d_1^*$ in Example~\ref{example2} is a PBIB
design based on the group divisible association scheme as in
Example~\ref{gdexample}, with $v^*_1=af,\ b^*_1={a\choose 2}f^2,
k^*_1=2, \ r^*_1=(a-1)f, \  \lambda_1=0, \ \lambda_2=1$. As seen
there, such a $d^*_1$ exists for   all integers $a,f(\geq 2)$. Also,
let the $d_2^*$ in Example~\ref{example2} be a BIB design with
$v^*_2=2g+1,\ b^*_2=\frac{1}{3}(2g+1)g, k^*_2=3, \ r^*_2=g, \
 \lambda=1$. Such a BIB design corresponds to the Steiner's triple system and it is well known (cf. Kirkman (1847)) that it exists for every integer $g(\geq 3)$
satisfying $g= 0\;{\rm or}\; 1\;({\rm mod}\; 3)$. Note that the BIB
design in Example~\ref{BIBDexample} belongs to this class with
$g=4$.

In our construction (\ref{nodes}), now take $t=2$, with $d_1$ and
$d_2$ chosen as the dual designs of $d_1^*$ and $d_2^*$,
respectively. Then recalling Definition~\ref{dual},  the parameters
of $d_1$ are
\begin{eqnarray}\label{d1parametersex2} &v_1={a\choose 2}f^2, \, b_1=af, \, r_1=2, \, k_1=(a-1)f,\nonumber\\
&\theta_1(1)=f-1,\,\,\, \theta_2(1)=(a-1)f,
\phi^1_{1,2}(1)=0,\,\,\,\phi^1_{2,2}(1)=(a-1)f, \end{eqnarray} and
the parameters of $d_2$ are \begin{eqnarray}\label{d2parameters}
&v_2=\frac{1}{3}(2g+1)g,\;b_2=2g+1,\;r_2=3,\;k_2=g,\nonumber \\
&\theta_1(2)=0,\;\theta_2(2)=2g,\;\phi^2_{2,2}(2)=2g-1.\end{eqnarray}
The KPS obtained from $d_1$ and $d_2$ via (\ref{nodes}) has
$v=v_1+v_2={a\choose 2}f^2+\frac{1}{3}(2g+1)g$ keys and
$n=b_1b_2=af(2g+1)$ nodes, there being $k=k_1+k_2=(a-1)f+g$ keys in
every node. For this KPS, substitution of  (\ref{d1parametersex2})
and (\ref{d2parameters})  in (\ref{muex2}) yields
$\mu_{12}=2+(a-1)f(2g+1)$ and hence from (\ref{pr1ex2}) and
(\ref{pr2ex2}) we get
\begin{eqnarray}\label{pr1pr2ex2}
{\rm Pr}_1&=&\frac{af+2g-1+2(a-1)fg}{af(2g+1)-1}, \nonumber \\
{\rm Pr}_2&\approx
&\frac{2(f-1)g}{af(2g+1)-1}\left[1-\left(1-\frac{\mu_{12}}{af(2g+1)-2}\right)^\eta
\right]. \nonumber \end{eqnarray}Similarly,  substitution of
(\ref{d1parametersex2}) and (\ref{d2parameters}) in (\ref{nex2}) and
(\ref{xiex2})  yields
\begin{eqnarray}
n_{02}&=&2g,\,\,n_{10}=f-1,\,\,n_{20}=(a-1)f,\,\,n_{22}=2(a-1)fg,\nonumber \\
\xi_{02}&=&\left\{1-\left(1-\frac{1}{af}\right)^s\right\}\left\{1-\left(1-\frac{3}{2g+1}\right)^s\right\},\nonumber \\
\xi_{10}&=&1-\left(1-\frac{1}{2g+1}\right)^s,\nonumber \\
\xi_{20}&=&\left\{1-\left(1-\frac{2}{af}\right)^s\right\}\left\{1-\left(1-\frac{1}{2g+1}\right)^s\right\},\nonumber \\
\xi_{22}&=&\left\{1-\left(1-\frac{2}{af}\right)^s\right\}\left\{1-\left(1-\frac{3}{2g+1}\right)^s\right\}.\nonumber
 \end{eqnarray}Theorem~\ref{fails} can now be easily used
to find ${\rm fail}(s)$.

On varying the values of $a, \ f$ and $g$ we can get various choices
of $d^*_1$ and $d^*_2$, leading to KPSs for a variety  of parameter
values. Two illustrative examples follow.

\begin{example}\label{ex6} Let  $a=2, \; f =21, \; g = 25$. Then for the resulting KPS, we have $v=866,$ $n = 2142,\,k = 46$, while the values of  ${\rm Pr}_1, \, {\rm Pr}_2,$ \, ${\rm Pr}={\rm Pr}_1+{\rm Pr}_2$ for various $\eta$ and the values of ${\rm fail}(s)$ for various $s$ are as: \vskip.1in
\begin{tabular}{|ccccccccc|}
\hline
$\eta $&1&2&3&4&5&10&15&20\\
\hline
${\rm Pr}_1$  & 0.5329 &0.5329& 0.5329 & 0.5329 & 0.5329 & 0.5329 & 0.5329 & 0.5329\\
  ${\rm Pr}_2$& 0.2342 &0.3510& 0.4092 & 0.4382 & 0.4527 & 0.4667 & 0.4671 & 0.4671 \\
            Pr& 0.7671 &0.8839& 0.9421 & 0.9711 & 0.9856 & 0.9996 & 1.0000 & 1.0000\\
\hline \hline
$s$&1&2&3&4&5&6&8&10\\
\hline
fail$(s)$   &0.0021&0.0089&0.0198&0.0340& 0.0510&0.0703&0.1141&0.1624\\
\hline
\end{tabular}
\end{example}

\begin{example}\label{ex7} Let $a = 2, \; f = 23, \; g = 22$. The resulting KPS has $v = 859$, $n = 2070, \, k= 45$ and the values of  ${\rm Pr}_1, \, {\rm Pr}_2,$ \, ${\rm Pr}={\rm Pr}_1+{\rm Pr}_2$ and ${\rm fail}(s)$ are as: \vskip.1in
\begin{tabular}{|ccccccccc|}
\hline
$\eta $&1&2&3&4&5&10&15&20\\
\hline
${\rm Pr}_1$  & 0.5321 &0.5321& 0.5321 & 0.5321 & 0.5321 & 0.5321 & 0.5321 & 0.5321\\
  ${\rm Pr}_2$& 0.2346 &0.3516& 0.4099 & 0.4390 & 0.4535 & 0.4675 & 0.4679 & 0.4679 \\
Pr            & 0.7667 &0.8837& 0.9420 & 0.9711 & 0.9856 & 0.9996 & 1.0000 &1.0000\\
\hline \hline
$s$&1&2&3&4&5&6&8&10\\
\hline
fail$(s)$&0.0022&0.0093&0.0206&0.0352&0.0527 &0.0724&0.1169&0.1658\\
\hline
\end{tabular}
\end{example}

\subsection {Use of a PBIB design based on the triangular association scheme and a BIB design}\label{triangular}

Now suppose the design $d_1^*$ in Example~\ref{example2} is a
triangular PBIB design as constructed in
Example~\ref{triangularexample}. Thus $d_1^*$ has  $v^*_1={m\choose
2}, \ b^*_1=3{m\choose 4}, \ k^*_1=2, \ r^*_1={{m-2}\choose 2}, \
\lambda_1=0, \ \lambda_2=1$, and as seen there, such a $d_1^*$
exists for every integer $m(\geq 4)$.
 Also, let us continue with $d_2^*$ as the BIB design considered in Subsection~\ref{gd}.

In our construction (\ref{nodes}), take $t= 2$, with $d_1$ and $d_2$
chosen as the dual designs of $d_1^*$ and $d_2^*$, respectively.
Then recalling Definition~\ref{dual}, the parameters of $d_1$ are
\begin{eqnarray}\label{d1parameters}
&v_1=3{m\choose 4},  \;b_1={m\choose 2},\;r_1=2, \;k_1={{m-2}\choose 2},\nonumber \\
&\theta_1(1)=2(m-2),\;\theta_2(1)={m-2\choose 2},\;
\phi^1_{1,2}(1)=m-3,\;\phi^1_{2,2}(1)={m-3\choose 2},\end{eqnarray}
while the parameters of $d_2$ are  as in (\ref{d2parameters}). The
KPS obtained from $d_1$ and $d_2$ via (\ref{nodes}) has
$v=3{m\choose 4}+\frac{1}{3}(2g+1)g$  keys and $n={m\choose
2}(2g+1)$ nodes, there being $k={{m-2}\choose 2}+g$ keys in every
node. For this KPS, substitution of  (\ref{d2parameters}) and
(\ref{d1parameters})
 in (\ref{muex2}) yields
$\mu_{12}=2(m-2)+{{m-3}\choose 2}(2g+1)$ and hence from
(\ref{pr1ex2}) and (\ref{pr2ex2})
\begin{eqnarray}
{\rm Pr}_1&=&\frac{m(m-1)+4g-2+2(m-2)(m-3)g}{m(m-1)(2g+1)-2}, \nonumber \\
{\rm Pr}_2&\approx
&\frac{8(m-2)g}{m(m-1)(2g+1)-2}\left[1-\left(1-\frac{2\mu_{12}}{m(m-1)(2g+1)-4}\right)^\eta\right].\
\nonumber\end{eqnarray}
 Similarly, substitution of (\ref{d2parameters}) and (\ref{d1parameters})
 in (\ref{nex2}) and (\ref{xiex2})  yields
\begin{eqnarray}
n_{02}&=&2g,\;n_{10}=2(m-2),\; n_{20}={m-2\choose 2}, \;n_{22}=(m-2)(m-3)g, \nonumber \\
\xi_{02}&=&\left\{1-\left(1-\frac{2}{m(m-1)}\right)^s\right\}\left\{1-\left(1-\frac{3}{2g+1}\right)^s\right\},\nonumber\\
\xi_{10}&=&1-\left(1-\frac{1}{2g+1}\right)^s,\nonumber\\
\xi_{20}&=&\left\{1-\left(1-\frac{4}{m(m-1)}\right)^s\right\}\left\{1-\left(1-\frac{1}{2g+1}\right)^s\right\},\nonumber\\
\xi_{22}&=&\left\{1-\left(1-\frac{4}{m(m-1)}\right)^s\right\}\left\{1-\left(1-\frac{3}{2g+1}\right)^s\right\}.
\nonumber \end{eqnarray}
 Theorem~\ref{fails} can now be employed
to find ${\rm fail}(s)$. Again, on varying $m$ and $g$ we can get
KPSs for a variety  of parameter values. Two illustrative examples
follow. \vskip.3in

 \begin{example}\label{ex4} Let $m = 9$ and $g =27$. The resulting KPS has $v=873,\,n = 1980,\,k = 48$ and the values of  ${\rm Pr}_1, \, {\rm Pr}_2,$ \, ${\rm Pr}={\rm Pr}_1+{\rm Pr}_2$ and ${\rm fail}(s)$ are as: \vskip.1in
\begin{tabular}{|ccccccccc|}
\hline
$\eta $&1&2&3&4&5&10&15&20\\
\hline
${\rm Pr}_1$  & 0.6180 &0.6180& 0.6180 & 0.6180 & 0.6180 & 0.6180 & 0.6180 & 0.6180\\
  ${\rm Pr}_2$& 0.1620 &0.2553& 0.3091 & 0.3400 & 0.3578 & 0.3805 & 0.3819 & 0.3820 \\
 ${\rm Pr}$   & 0.7800 &0.8733& 0.9271 & 0.9580 & 0.9758 & 0.9985 & 0.9999 &1.0000\\
\hline \hline
$s$&1&2&3&4&5&6&8&10\\
\hline
fail$(s)$&0.0021&0.0094&0.0210&0.0362& 0.0544 &0.0750&0.1216&0.1728\\
\hline
\end{tabular}
\end{example}

\begin{example}\label{ex5} Let $m = 8$ and $g = 31$. The resulting KPS has $v=861,\,n=1764, \,k=46$ and the values of  ${\rm Pr}_1, \, {\rm Pr}_2,$ \, ${\rm Pr}={\rm Pr}_1+{\rm Pr}_2$ and ${\rm fail}(s)$ are as: \vskip.1in \noindent
\begin{tabular}{|ccccccccc|}
\hline
$\eta $&1&2&3&4&5&10&15&20\\
\hline
${\rm Pr}_1$  & 0.5780 &0.5780& 0.5780 & 0.5780 & 0.5780 & 0.5780 & 0.5780 & 0.5780\\
  ${\rm Pr}_2$& 0.1538 &0.2515& 0.3136 & 0.3531 & 0.3782 & 0.4175 & 0.4215 & 0.4220 \\
    Pr        & 0.7318  &0.8295& 0.8916 & 0.9311 & 0.9562 & 0.9955 & 0.9995  &1.0000\\
\hline \hline
$s$&1&2&3&4&5&6&8&10\\
\hline
fail$(s)$&0.0023&0.0103&0.0230&0.0396& 0.0593&0.0815&0.1312&0.1853\\
\hline
\end{tabular}
\end{example}

\subsection{Use of a PBIB design based on the Latin square type association scheme and a BIB design}\label{transversal}

Now suppose the design $d_1^*$ in Example~\ref{example2} is a PBIB
design based on the Latin square type association scheme and having
parameters $v^*_1=p^2, \ b^*_1=\tilde{k}p, \ k^*_1= p, \
r^*_1=\tilde{k}, \ \lambda_1=0, \ \lambda_2=1$. Such a design exists
when $p(\geq 3)$ and $\tilde{k} (< p+1)$ are such that $\tilde{k}-2$
mutually orthogonal Latin squares of order $p$ are available, cf.
Definition~\ref{Latinsquarescheme}. Hence following
 Definition~\ref{dual}, its dual design  $d_1$ has parameters
 \begin{eqnarray}\label{d1parametersex3}
 &v_1=\tilde{k}p, \, \; b_1 = p^2, \, \; r_1= p, \,\; k_1= \tilde{k}, \ \; \theta_1(1) = (p-1)(p+1-\tilde{k}), \, \; \theta_2(1)=
 \tilde{k}(p-1), \nonumber \\
 &\phi^1_{1,2}(1)=\tilde{k}(p-\tilde{k}), \,
 \phi^1_{2,2}(1)=\tilde{k}(\tilde{k}-1).\end{eqnarray}
We continue with $d_2$ as in the last two subsections and
(\ref{d2parameters}) continues to hold for $d_2$.   In our
construction (\ref{nodes}), now take $t = 2$, with $d_1$ and $d_2$
chosen as above.

Clearly, the KPS obtained from $d_1$ and $d_2$ via (\ref{nodes}) has
$v = \tilde{k}p + \frac{1}{3}(2g+1)g$ keys and $n = p^2(2g+1)$
nodes, there being $k = \tilde{k}+g $ keys in every node. For this
KPS, substitution of (\ref{d2parameters}) and
(\ref{d1parametersex3}) in
 (\ref{muex2}) yields $\mu_{12}=2+2\tilde{k}(p-\tilde{k})+\tilde{k}(\tilde{k}-1)(2g+1)$ and hence from
(\ref{pr1ex2}) and (\ref{pr2ex2}) we get
\begin{eqnarray}\label{pr1pr2ex3}
{\rm Pr}_1&=&\frac{p^2+2g-1+2\tilde{k}(p-1)g}{p^2(2g+1)-1}, \nonumber\\
{\rm Pr}_2&\approx
&\frac{2(p-1)(p+1-\tilde{k})g}{p^2(2g+1)-1}\left[1-\left(1-\frac{\mu_{12}}{p^2(2g+1)-2}\right)^\eta
\right]. \nonumber\end{eqnarray}Similarly,  substitution of
(\ref{d2parameters}) and (\ref{d1parametersex3}) in (\ref{nex2}) and
(\ref{xiex2})  yields
\begin{eqnarray}
n_{02}&=&2g,\,\,n_{10}=(p-1)(p+1-\tilde{k}),\,\,n_{20}=\tilde{k}(p-1),\,\,n_{22}=2\tilde{k}(p-1)g,\nonumber \\
\xi_{02}&=&\left\{1-\left(1-\frac{1}{p^2}\right)^s\right\}\left\{1-\left(1-\frac{3}{2g+1}\right)^s\right\},\nonumber \\
\xi_{10}&=&1-\left(1-\frac{1}{2g+1}\right)^s,\nonumber \\
\xi_{20}&=&\left\{1-\left(1-\frac{1}{p}\right)^s\right\}\left\{1-\left(1-\frac{1}{2g+1}\right)^s\right\},\nonumber \\
\xi_{22}&=&\left\{1-\left(1-\frac{1}{p}\right)^s\right\}\left\{1-\left(1-\frac{3}{2g+1}\right)^s\right\}.\nonumber
 \end{eqnarray} Theorem~\ref{fails} can now be easily used
to find ${\rm fail}(s).$  Again, KPSs for a variety of parameter
values can  be obtained by varying the values of $p, \; \tilde{k}$
and $g$. Two illustrative examples follow.

\begin{example}\label{ex8}Let  $p=17, \; \tilde{k }=12, \; g = 28$. Then the resulting KPS has $v=736,$ $n = 16473,\,k = 40$ and the values of  ${\rm Pr}_1, \, {\rm Pr}_2,$ \, ${\rm Pr}={\rm Pr}_1+{\rm Pr}_2$ and ${\rm fail}(s)$ are as:\vskip.1in
\begin{tabular}{|ccccccccc|}
\hline
$\eta $&1&2&3&4&5&10&15&20\\
\hline
${\rm Pr}_1$  & 0.6736 &0.6736& 0.6736 & 0.6736 & 0.6736 & 0.6736 & 0.6736 & 0.6736\\
  ${\rm Pr}_2$& 0.1515 &0.2327& 0.2762 & 0.2995 & 0.3120 & 0.3258 & 0.3264 & 0.3264 \\
            Pr& 0.8251 &0.9063& 0.9498 & 0.9731 & 0.9856 & 0.9994 & 1.0000 & 1.0000\\
\hline \hline
$s$&1&2&3&4&5&6&8&10\\
\hline
fail$(s)$   &0.0030&0.0115 &0.0244 &0.0410 & 0.0606 &0.0826 &0.1320 &0.1857\\
\hline
\end{tabular}
\end{example}

 \begin{example}\label{ex9} Now let  $p=19, \; \tilde{k} =13, \; g = 28$. Then the resulting KPS has $v=779,$
 $n = 20577,\,k = 41$ and the values of  ${\rm Pr}_1, \, {\rm Pr}_2,$ \, ${\rm Pr}={\rm Pr}_1+{\rm Pr}_2$ and ${\rm fail}(s)$ are as: \vskip.1in
\begin{tabular}{|ccccccccc|}
\hline
$\eta $&1&2&3&4&5&10&15&20\\
\hline
${\rm Pr}_1$  & 0.6571 &0.6571& 0.6571 & 0.6571 & 0.6571 & 0.6571 & 0.6571 & 0.6571\\
  ${\rm Pr}_2$& 0.1508 &0.2353& 0.2826 & 0.3091 & 0.3240 & 0.3419 & 0.3428 & 0.3429 \\
            Pr& 0.8079 &0.8924& 0.9397 & 0.9662 & 0.9811 & 0.9990 & 0.9999 & 1.0000\\
\hline \hline
$s$&1&2&3&4&5&6&8&10\\
\hline
fail$(s)$   &0.0028&0.0104 &0.0221 &0.0372 & 0.0551 &0.0753 &0.1209 &0.1710\\
\hline
\end{tabular}
\end{example}

\section{ Shared key discovery}\label{shared}

A major advantage of our construction in (\ref{nodes}) is that it
makes the task of discovering the keys shared by any two nodes of
the resulting KPS quite straightforward. This happens because of the
following reasons:

(a) Consider any two distinct nodes $A$ and $A'$. From (\ref{nodes})
and Definition~\ref{projection} it is clear that $proj(A,i)$ and
$proj(A',i')$ do not have any common symbol whenever $i\neq i'$.
Hence, the set of keys (symbols) common to $A$ and $A'$ equals the
union of the sets of symbols common to $proj(A,i)$ and $proj(A',i)$,
the union being over all $i, 1\leq i \leq t.$ As a result, in order
to discover the keys shared by $A$ and $A'$, it suffices to find the
set of symbols common to $proj(A,i)$ and $proj(A',i)$,
\emph{separately} for each $i,$ $1\leq i \leq t.$ This is much
simpler than comparing the entire sets of keys in $A$ and $A'$.

(b) Turning now to the identification of the set of symbols common
to $proj(A,i)$ and $proj(A',i)$  for any $i$, from
Definition~\ref{projection} we see that this set is nothing but the
set of symbols common to two blocks of $d_i$. Therefore, in view of
the duality between $d_i$ and the design $d^*_i$ that we originally
started with, this set is simply the set of blocks labels where the
corresponding two symbols of $d^*_i$ occur together. Thus
identification of this set becomes particularly easy if the symbols
and blocks in $d^*_i$ can be properly labeled so as to obtain
algebraically a listing of the symbols appearing in each block of
$d^*_i$. Since the $d^*_i$ considered here are PBIB or BIB designs,
such labeling is possible under  wide generality. For instance, the
commonly used cyclic constructions of these designs, based on one or
more initial sets, readily allow such labeling. This kind of
labeling is also possible for the constructions described in
Examples~\ref{gdexample} and \ref{triangularexample}.

Indeed in  construction (\ref{nodes}), each $d^*_i$ can potentially
be \emph{any} PBIB design with $\lambda_1=0, \lambda_2=1$ or\emph{
any } BIB design with $\lambda=1$. Because of such diversity, it is
unrealistic in the limited space of this paper to attempt to give an
account of the labeling of blocks and symbols, mentioned in (b)
above, encompassing \emph{all} possibilities for $d^*_i$, $i=1,
\ldots, t$. For illustration, therefore,  we now revisit the setup
of Subsection~\ref{gd} in some detail; those of
Subsections~\ref{triangular} and \ref{transversal} are  briefly
touched upon later.

Recall that in Subsection~\ref{gd}, $d^*_1$ is a group divisible
PBIB design constructed as in Example~\ref{gdexample}. Also $d^*_2$
is a BIB design belonging to the Steiner's triple system, and as
seen below, it is generated via a cyclic construction. The
parameters of these designs are as described in Subsection~\ref{gd}.
The facts noted below in (A) and (B) for these two designs will be
useful.

 \textbf{(A)} \emph{Labels for symbols and blocks of
$d^*_1$:} Denote the $af$ symbols of $d^*_1$ by ordered pairs
$\beta\gamma$, where $\beta\gamma$ is the $\gamma$th symbol of the
$\beta$th group; $1\leq \beta \leq a$ and $1\leq \gamma \leq f$.
Then as indicated in Example~\ref{gdexample}, its ${a\choose 2}f^2$
blocks are $\{\beta\gamma, \tilde{\beta}\delta\}, $ and let these be
labeled as $\beta\tilde{\beta}\gamma\delta$, say, where $1\leq \beta
< \tilde{\beta}\leq a$ and $\gamma, \delta \in \{1, 2, \ldots, f
\}$. Thus, any two distinct symbols $\beta\gamma$ and
$\tilde{\beta}\delta$ occur together in some block \emph{if and only
if} $\beta\neq \tilde{\beta}$, and if this happens then the unique
block where they occur together has label
$\beta\tilde{\beta}\gamma\delta$ if $\beta < \tilde{\beta}$ or
$\tilde{\beta}\beta\delta\gamma$ if $\tilde{\beta} < \beta $. Let
 the label for this block be identified as
$L_1(\beta\gamma,\tilde{\beta}\delta)$.

Similarly, the $(a-1)f$ blocks where any symbol $\beta\gamma$ occurs
have labels  (i) $\beta\tilde{\beta}\gamma\delta$ , where  $ \beta <
\tilde{\beta}\leq a$ and $\delta \in\{1, 2, \ldots, f\}$, and (ii)
$\tilde{\beta}\beta\delta\gamma$ where $1\leq \tilde{\beta}< \beta$
and $\delta\in \{1, 2,\ldots, f\}$. Let $V_1(\beta\gamma)$ be the
collection of these $(a-1)f$ block labels. \qed

 \textbf{(B)} \emph{Labels for symbols and blocks of
$d^*_2$:} Let $g = 1$ mod $3$ in $d^*_2$, i.e., $g=3h+1$ for some
integer $h(\geq 1)$. So $d^*_2$ involves $6h+3$ symbols and
$(2h+1)(3h+1)$ blocks. Denote these symbols of $d^*_2$ by $\zeta_u$
where  $\zeta \in\{0, 1,\ldots, 2h\}, \, u=0,1,2.$ Then,
%essentially following Dey (2010),
the blocks of $d^*_2$ can be represented and labeled as
 $$ \{(y+z)_x , \, (z-y)_x, \, z_{x+1}\} = xyz, \, \,  {\rm say, \, and } \,
 \,
 \{z_0, \, z_1, \, z_2 \} = 0z, \, {\rm say},$$ where $x, y$ and $z$ range
over $\{0,1,2\}, \, \{1, \ldots, h\} $ and $\{0, 1, \ldots, 2h \}$,
respectively, and the subscript $x+1$ is reduced modulo 3, while
$y+z$ and $z-y$ are reduced modulo $2h+1$. There is a unique block
where two distinct symbols $\zeta_u$ and $\tilde{\zeta}_w, \,
(\zeta, u) \neq (\tilde{\zeta}, w)$, occur together and let the
label for this block be identified as $L_2(\zeta_u,
\tilde{\zeta}_w)$.

 Since $y$ ranges over $\{1,\ldots, h\}$, the
following are not hard to observe:

% [verified from first principles for h = 2]
\noindent (a) Let $u=w$ and $\zeta \neq \tilde{\zeta}$. Then
$L_2(\zeta_u, \tilde{\zeta}_u) = uyz$, where $ z = (\zeta +
\tilde{\zeta})/2$ mod $ 2h+1$ and $y=(\zeta - \tilde{\zeta})/2$  or
$(\tilde{\zeta} -\zeta)/2$ mod $2h+1$, depending on whether $(\zeta
- \tilde{\zeta})/2$ mod
 $2h+1$ belongs to $\{1, \ldots, h\}$ or $\{h+1,\ldots, 2h\}$.

\noindent (b) Let $u\neq w$  and $\zeta = \tilde{\zeta}$. Then
$L_2(\zeta_u, \zeta_w) =0\zeta$.

\noindent (c) Let $u\neq w$  and $\zeta \neq \tilde{\zeta}$.  Then
$L_2(\zeta_u, \tilde{\zeta}_w) = xyz$, where $(x, z) = (u,
\tilde{\zeta})$ or $(w, \zeta)$, depending on whether $w = u+1$ or
$u = w+1$ mod 3 and $y= \zeta - \tilde{\zeta}$ or $\tilde{\zeta} -
\zeta $ mod $2h+1$, depending on whether $\zeta - \tilde{\zeta}$ mod
$2h+1$  belongs to $\{1, \ldots, h\}$ or $\{h+1,\ldots, 2h\}$.

Similarly, the $g(=3h+1)$ blocks where any symbol $\zeta_u$ occurs
are labeled as (i) $uyz$, where $y \in \{1,\ldots, h\}$ and  $z =
\zeta \pm y$   mod $2h+1$, (ii) $(u-1)y\zeta$, where $y \in
\{1,\ldots, h\}$ and $u-1$ is reduced mod 3,  and (iii) $0\zeta$.
Let $V_2(\zeta_u)$  be the collection of these $3h+1$ block
labels.\qed

Returning to the setup of Subsection~\ref{gd}, consider now the KPS
constructed as in (\ref{nodes}), with $t = 2$ and $d_1$ and $d_2$
chosen as the dual designs of $d^*_1$ and $d^*_2$, respectively,
where $d^*_1$ and $d^*_2$  are as detailed in the facts (A) and (B)
above. As seen in Subsection~\ref{gd}, this KPS has $v={a\choose
2}f^2+\frac{1}{3}(2g+1)g = {a\choose 2}f^2+ (2h+1)(3h+1)$ keys and $
n = af(6h+3)$ nodes. Since $d_1$ and $d_2$  are obtained by
interchanging the roles of symbols and blocks in $d^*_1$ and
$d^*_2$, respectively, it is clear from (\ref{nodes}) that the $v$
keys correspond to the block labels of $d^*_1$ and $d^*_2$, while
the $n$ nodes correspond to ordered pairs whose first member is a
symbol of $d^*_1$ and second member is a symbol of $d^*_2$.

Thus, using the facts in (A) and (B), the $v$ keys can be denoted by
$\beta\tilde{\beta}\gamma\delta, \, xyz$ and $0z$, where $1\leq
\beta < \tilde{\beta }\leq a$ and $\gamma, \delta \in \{ 1,2,\ldots,
f\}$, while $x, y$ and $z$ range over $\{0, 1, 2\}$, $\{ 1, \ldots,
h\}$ and $\{0, 1, \ldots, 2h\}$, respectively. Similarly, the $n$
nodes can be labeled as $(\beta\gamma, \zeta_u)$, where $1\leq \beta
\leq a, \, 1\leq \gamma \leq f$, and $u$ and $\zeta$ range over
$\{0, 1, 2\}$ and $\{0, 1,\ldots, 2h\}$, respectively. Then clearly,
the keys appearing in any node $(\beta\gamma, \zeta_u)$ are given by
the labels of the blocks of $d^*_1$ containing the symbol
$\beta\gamma$ and the labels of the blocks of $d^*_2$ containing the
symbol $\zeta_u$. Hence, as discussed in the beginning of this
section, the keys shared by two distinct nodes $(\beta\gamma,
\zeta_u)$, and $(\tilde{\beta}\delta, \tilde{\zeta}_w)$ are given by
the labels of the blocks of $d^*_1$ containing both $\beta\gamma $
and $\tilde{\beta}\delta$ and the labels of the blocks of $d^*_2$
containing both $\zeta_u$ and $\tilde{\zeta}_w$, i.e., using the
facts noted in (A) and (B), these shared keys are as described
below:

\noindent (i) the keys in $V_1(\beta\gamma)$ and key $L_2(\zeta_u,
\tilde{\zeta}_w)$, if $\beta\gamma = \tilde{\beta}\delta$ and
$(\zeta,u) \neq(\tilde{\zeta},w)$;

\noindent (ii) the keys in $V_2(\zeta_u),$  if $\beta=\tilde{\beta},
\, \gamma \neq \delta$ and
 $(\zeta,u)=(\tilde{\zeta},w)$;

\noindent (iii) the key $L_1(\beta\gamma, \tilde{\beta}\delta)$ and
the keys in $V_2(\zeta_u),$  if $\beta \neq \tilde{\beta} $ and
$(\zeta,u)= (\tilde{\zeta},w)$;

\noindent (iv) the key $L_2(\zeta_u, \zeta_w)$  if
$\beta=\tilde{\beta}, \, \gamma \neq \delta$, and $(\zeta,u) \neq
(\tilde{\zeta},w)$;

\noindent (v) the keys $L_1(\beta\gamma, \tilde{\beta}\delta)$ and
$L_2(\zeta_u, \tilde{\zeta}_w)$ if $\beta \neq \tilde{\beta}$ and
$(\zeta,u) \neq (\tilde{\zeta},w)$

Thus the keys shared by any two distinct nodes can be found readily
from the node labels. Consider any two nodes $A$ and $A'$  in each
other's neighborhood and by our construction  as described above,
suppose they are assigned labels $(\beta\gamma, \zeta_u)$ and
$(\tilde{\beta}\delta, \tilde{\zeta}_w)$, respectively. In the
shared-key discovery phase, node $A$ only broadcasts the four values
$\beta, \gamma, \zeta$ and $u$. Once node $A'$ receives these four
values, it simply checks them against the corresponding four values
in its own label, decides on one of the five cases in (i)-(v) above
and accordingly, it immediately identifies its common keys with $A$.
Thus there is no need to solve any equations nor any complicated
computations are involved. Path-key establishment is also similarly
straightforward.

For further illustration, we revisit the second example of
Subsection~\ref{gd}, where $a = 2, \, f =23$ and $g = 22$. Then, as
discussed above, the keys of the resulting KPS can be denoted by
$12\gamma\delta, \, xyz$ and $0z$, where $\gamma, \delta \in \{1, 2,
\ldots, 23\}$, while $x, y$ and $z$ range over $\{0, 1, 2\}$,
$\{1,\ldots, 7\}$ and $\{0, 1,\ldots,  14\}$, respectively.
Similarly, the nodes of this KPS can be labeled as $(\beta\gamma,
\zeta_u)$ where $\beta= 1$ or $2,$ $1\leq \gamma \leq 23,$ and $u$
and $\zeta$ range over $\{0, 1, 2\}$ and \{0, 1,\ldots, 14\},
respectively. From (i) above, the keys shared, for example, by the
nodes $(16, 4_0)$ and $(16, 6_0)$ are $126\delta$, $1\leq \delta\leq
23,$ which constitute $V_1(16),$ and $L_2(4_0, 6_0)=015$. Similarly,
from (v) above, the nodes $(22, 5_1)$ and $(13, 6_2)$ share the keys
$L_1(22,13)   = 1232$ and $L_2(5_1, 6_2)=116.$

The other applications considered in Section~\ref{applications}
allow equally simple discovery of shared keys. The symbols and
blocks of the triangular PBIB design in Subsection~\ref{triangular}
can be represented along the lines of (A) above. Also, following Lee
and Stinson (2008), the blocks of $d_1$ in
Subsection~\ref{transversal} can be so labeled that one can readily
identify the common symbol, if any, between two given blocks.
Furthermore, if $g = 0$ mod $3$ for the BIB design $d^*_2$, then one
can represent its symbols and blocks in a manner similar to (B)
above. These representations readily yield the counterparts of $V_1,
\,V_2, \, L_1$ and $L_2$  for these designs. As a result, for
constructions involving these designs, keys shared by any two
distinct nodes can again be found easily from the node labels.

\section{Comparison of our method with some existing ones}\label{comparison}

In this paper, we have given a general method for construction of
KPSs using duals $d_1, \ldots, d_t$ of PBIB or BIB designs. The most
important features of our method can be summarized as follows:

\noindent (i) It is applicable to any prespecified intersection
threshold $q \geq 1$.

\noindent (ii) It allows the construction of KPSs for a wide
spectrum of parameter values, namely, the number of nodes $n$, the
number of keys per node $k$ and the key pool size $v$, thus enabling
the user to find a suitable KPS in a given context.

\noindent (iii) It ensures that $n$ is multiplicative in the numbers
of blocks of $d_1, \ldots, d_t$ while $k$ is additive in the block
sizes of these designs. This allows a large $n$ and, at the same
time, keeps $k$ in check.

\noindent (iv) It comes along with explicit formulae for the local
connectivity and resiliency metrics as given by Pr and fail$(s)$. It
also keeps the tasks of shared key discovery and path key
establishment simple.

\noindent As seen earlier, for instance, in the beginning of
Section~\ref{applications} and in Remarks~\ref{remark},
\ref{algebraic}, because of (i)-(iv) above, our method has several
advantages compared to the existing ones. We now indicate these
advantages in some more detail.

First note that in contrast to (i), the existing methods based on
combinatorial designs are typically meant for specific values of
$q$, such as $q=1$ in Camtepe and Yener (2004, 2007), Lee and
Stinson (2005a), Chakrabarty et al. (2006), Dong et al. (2008),  Ruj
and Roy (2007) and Ruj et al. (2009), or separately for $q=1$ and
$q=2$ in Lee and Stinson (2008).

Next, as a consequence of (ii), our method allows us to obtain KPSs
 for networks where the number of
nodes $n$ need not be of any specialized form, such as the forms
$p(p-1)/2$ or $p(p-1)$ as in Ruj and Roy (2007), or  the forms $p^2$
(for $q=1$) or $p^3$ (for $q=2$), $p$ a prime/prime power, as in Lee
and Stinson (2008). Furthermore, because of (iii) and (iv), this can
be achieved with a control on the number of keys $k$ per node, while
assuring  good values of the performance metrics. To understand why
this is important, let $q=2$ and suppose we start with a scheme of
Lee and Stinson (2008) with $n$ equal to the lowest prime power of
the form $p^3$ that exceeds the target number of nodes. If we then
discard the unnecessary node allocations to get the final scheme for
use, this final scheme will not preserve the Pr and fail$(s)$ values
of the original scheme and hence the properties of the final scheme
in this regard can become quite erratic. This is because, these
performance metrics of the original scheme depend on the pattern of
the keys allocated to the different nodes, this allocation having
been done by exploiting the structure of some combinatorial design,
and once a large number of the allocated nodes are discarded, the
underlying combinatorial structure is disrupted, leading to a scheme
with uncertain local connectivity and resiliency properties.

For illustration, suppose it is desired to obtain a KPS with about 16500 nodes.
Then our Example~\ref{ex8} gives a scheme with 16473 nodes with demonstrated good values of the performance metrics.
The closest higher prime power of the form $p^3$ is $27^3=19683.$
If we start with the scheme of Lee and Stinson (2008) with allocation for 19683 nodes,
 we will have to delete the allocation for about (19683-16500=)3183 nodes constituting 16.17\%
 of the original 19683 nodes. After such large scale deletion, the Pr and fail$(s)$ values of the final
 scheme very much depend on the particular nodes deleted and hence become quite arbitrary. Similarly,
 if about 20500 nodes are needed, then our Example~\ref{ex9} gives a scheme with 20577 nodes and assured properties while
 the nearest scheme of Lee and Stinson (2008) with $29^3=24389$ nodes entails a deletion of about 3889, i.e., 15.95\%,
 of the nodes, leading to unpredictable performance. In either of these situations, the constructions in Ruj and Roy (2007),
  with  $n=p(p-1)/2$ or $p(p-1)$ and $k =2(p-2)$, can bring $n$ close to the target but at the cost of prohibitively large
  (i.e., 250 or even larger) values of $k$. In contrast, the schemes in our Examples~\ref{ex8} and \ref{ex9} involve only
  40 and 41 keys per node. The additive nature of $k$ in our construction, as mentioned in (iii) above, helps in achieving this.

Finally, as noted in (iv), our method comes along with explicit and
readily applicable formulae for ${\rm Pr}= {\rm Pr}_1 + {\rm Pr}_2$
and fail$(s)$, and also keeps the tasks of shared key discovery and
path key establishment simple. Not all of these aspects have been
explored in many of the existing constructions of KPSs via
combinatorial designs, and even when this is done, analytical
results on Pr and fail$(s)$ are not always available. For example,
 Dong et al. (2008) studied only ${\rm Pr}_1$ and fail$(1)$ for their scheme. Again,  as seen in
Remark~\ref{algebraic}, the quantity ${\rm Pr}_2$ in the Lee and
Stinson (2008) scheme for $q=2$ does not admit an explicit
expression and its calculation calls for design specific numerical
enumeration which can be difficult when the number of nodes is
large. Similarly, Ruj and Roy (2007) and Ruj et al. (2009) gave some
bounds on the expected number of links that will be broken if a
specified number of nodes are compromised in their schemes and
reported associated simulation results, but did not study fail$(s)$.
 Incidentally, their schemes have ${\rm Pr}_1=1$, a feature shared
also by our construction when the  initial designs $d^*_1, \ldots
,d^*_t$ are all taken as BIB designs with $\lambda=1$; cf.
Example~\ref{example1}. However, as argued in Lee and Stinson
(2008), a scheme with ${\rm Pr}_1=1$ will have poor connectivity in
the event of node compromise as reflected in large fails$(s)$
values. This is why we have focused on schemes with good values of
Pr rather than attempting to have ${\rm Pr}_1=1$.

To sum up, our method of construction is a broad spectrum one which
supplements and improves upon the existing methods from various
considerations. It is applicable to any intersection threshold $q
\geq 1$ and allows the construction of KPSs for widely diverse
parameter values. The fact that it is supported by a detailed study
of the performance metrics, including explicit formulae for Pr and
fail$(s)$, further enhances the scope of its application.

\section*{Acknowledgement}
The authors thank two referees for their insightful comments which
led to an enhancement of the contents and presentation in this
version.  The work of AD was supported by the Indian National
Science Academy under the Senior Scientist Scheme of the Academy.
The work of RM was supported by the J. C. Bose National Fellowship
of the Govt. of India and a grant from the Indian Institute of
Management Calcutta.

\end{document}